\documentclass[journal]{IEEEtran}
\usepackage{amsmath, amssymb, amsfonts}
\usepackage{algorithmic}
\usepackage{algorithm}
\usepackage{array}
\usepackage[caption=false,font=scriptsize,labelfont=sf,textfont=sf]{subfig}
\usepackage{textcomp}
\usepackage{stfloats}
\usepackage{url}
\usepackage{verbatim}
\usepackage{cite}
\usepackage{booktabs}

\usepackage{tabularx, multirow,multicol,colortbl,array}
\usepackage{graphicx, color, array}
\usepackage[dvipsnames]{xcolor}
\usepackage{float}
\usepackage{comment,psfrag}
\usepackage{enumitem}
\usepackage{tikz,tikzscale,bigints,pgfplots}
\pgfplotsset{compat=1.18}
\usetikzlibrary{positioning, arrows.meta, colorbrewer, calc, fit, matrix, external}
\usepgfplotslibrary{fillbetween}
\usepackage{multirow}

\pgfplotsset{
  colormap={blacktoblacktenwhite}{
    rgb255(0cm)=(255,140,0);   % Orange
    rgb255(1cm)=(0,0,0)    % Black
  }
}

\newif\iffigureson
\figuresontrue 
\begin{document}
% \title{Statistical Analysis and Practical Receiver Design via Block EM Algorithms for OFDM Systems Under Bursty Impulsive Noise}
% \title{Statistical Analysis and Block EM-Based Detection for OFDM Systems in Bursty Impulsive Noise}
% \title{On the Frequency-Domain Statistics and Block-EM Based Estimation for OFDM Systems in Bursty Impulsive Noise}
\title{Statistical Characterization and Block-EM Estimation of Frequency-Domain NSI for OFDM Systems in Bursty Impulsive Noise}
\author{Chin-Hung Chen,~\IEEEmembership{Graduate Student Member,~IEEE,}
Wim van Houtum,~\IEEEmembership{Senior Member,~IEEE,}
\\Yan Wu, 
and Alex Alvarado,~\IEEEmembership{Senior Member,~IEEE}
        % <-this % stops a space
\thanks{Manuscript received xx, xxxx; revised xx, xxxx.}% <-this % stops a space
\thanks{C.-H. Chen, W. van Houtum, and A. Alvarado are with the Electrical Engineering Department, Eindhoven University of Technology, 5600MB Eindhoven, The Netherlands. W. van Houtum is also with NXP Semiconductors, High Tech Campus 60, 5656AE Eindhoven, The Netherlands.  \textit{(e-mails: c.h.chen@tue.nl;  w.j.v.houtum@tue.nl;  a.alvarado@tue.nl)}.} 
\thanks{Y. Wu is with NXP Semiconductors, High Tech Campus 60, 5656AE Eindhoven, The Netherlands. \textit{(e-mail: yan.wu\_2@nxp.com)}.}% <-this % stops a space

}

% The paper headers
\markboth{}%
{Shell \MakeLowercase{\textit{et al.}}: A Sample Article Using IEEEtran.cls for IEEE Journals}

\maketitle

\begin{abstract}
Impulsive noise (IN), characterized by its high power and non-Gaussian distribution, poses a critical challenge in modern orthogonal frequency-division multiplexing (OFDM) systems, driven by the proliferation of electronic devices. Current IN mitigation techniques rely heavily on time-domain processing. These methods are applied before the discrete Fourier transform (DFT), introducing additional complexity, failing to align with OFDM's inherent frequency-domain processing flow, and risking the destruction of subcarrier orthogonality due to imperfect IN subtraction. To address these limitations, we propose a frequency-domain, block-based framework for mitigating IN. The statistical representation of IN in the frequency domain is first derived using a transformed Gaussian mixture model. Based on this model, we develop an optimal receiver that leverages perfect noise state information (NSI), thereby identifying scenarios in which NSI is critical. We then propose an unsupervised block-based expectation–maximization (EM) framework for NSI estimation and develop three variants for evaluation. These include a simple symbol-by-symbol variance-updated EM, a sequence-based transition-updated EM, and a MAP-based EM that exploits a sparsity-promoting prior to automatically prune the number of states. Our frequency-domain design operates after the DFT, seamlessly integrates with the OFDM processing chain, preserves subcarrier orthogonality, and leverages the known IN block structure to achieve substantial performance gains without the immense complexity of time-domain impulse reconstruction.

\end{abstract}

\section{Introduction}\label{sec:intro}
\IEEEPARstart{I}{mpulsive} noise (IN) exhibits high-power, non-Gaussian characteristics and has become a prominent impairment in modern communication systems due to the widespread proliferation of electronic devices such as switching gears, power lines, DC-DC converters, and other electronic equipment. This increasing prominence is evidenced by measurement studies across various environments, such as electric vehicles \cite{Maouloud21, Gao15, Pliakostathis20,chc24_1} and power substations \cite{Shan09, Sacuto14}. To design robust receivers, several IN models have been proposed based on empirically measured data. Among the best-known of these models, the Middleton Class A model \cite{Middleton97} is widely utilized because it provides a tractable probability density function (PDF) based on a Poisson distribution. However, the Middleton Class A model cannot capture the bursty temporal correlations observed in practical channels. To address this critical limitation, hidden Markov models (HMMs) have been adopted for IN modeling \cite{Zimmermann02, Dario09, MMA}, specifically to capture the correlation between adjacent impulsive samples.

For single-carrier communication systems, optimal maximum a-posteriori (MAP)-based receiver designs that exploit the IN statistics have been proposed for different IN models with different transmitter formats \cite{Dario09, MMA, Alam20, chc25_1}. To further enhance the receiver performance, powerful coding systems in combination with the IN detector have been extensively investigated, such as low-density parity-check (LDPC) codes \cite{MMA, Dario09, Alam20}, convolutional codes \cite{Mitra10, Mengistu14, chc25_1}, and turbo-like codes \cite{Tseng14, Umehara04}. Crucially, the performance of these MAP-based detector designs is fundamentally dependent on accurate knowledge of the statistics of IN, which has motivated extensive studies on the estimation of the noise state information (NSI). Estimation methods proposed in the literature include expectation-maximization (EM)-based parameter estimation \cite{Zabin91, Mitra09, Awino19, CYC21, chc25_2}, sampling techniques \cite{Hou18}, and neural-network-aided approaches \cite{Barka23}.

In modern communication systems, orthogonal frequency-division multiplexing (OFDM) has become a widely used modulation technique due to its ability to transform a frequency-selective channel into a set of parallel flat-fading subchannels, which greatly simplifies equalization. Consequently, OFDM is a core technology adopted in standards such as power line communications, digital audio broadcasting (DAB), and wireless local area networks (WLANs). However, the performance of conventional OFDM receivers is known to degrade severely in the presence of bursty IN. Comprehensive performance analyses regarding the impact of IN on OFDM systems can be found in \cite{Souissi23, Shongwe15, Suraweera04, Haring02}.

To mitigate the detrimental effects of IN, traditional receiver designs have focused extensively on time-domain processing. Techniques like adaptive clipping and blanking \cite{Ndo10, Oh17, Epple17} are favored for their straightforward implementation, but they are inherently suboptimal and prone to causing inter-carrier interference (ICI). Additionally, these techniques rely exclusively on signal processing before the discrete Fourier transform (DFT). This reliance forces them to operate before the OFDM demodulator, failing to exploit the intrinsic frequency-domain benefits.

Advanced signal reconstruction methods, such as compressed sensing \cite{Caire08, Lampe11, Liu16} and sparse Bayesian learning (SBL) \cite{Lin11, Lin13, Korki16}, aim to reconstruct sparse time-domain impulses and subsequently subtract them from the received signal, yielding near-optimal solutions. However, these iterative techniques, such as iterative sparse reconstruction of time-domain IN \cite{Korki16, Lin13} and iterative IN mitigation \cite{Chien15}, require complex processing and a large null-tone overhead to reconstruct the time-domain impulses. Moreover, time-domain reconstruction-based methods are susceptible to estimation errors that can disrupt subcarrier orthogonality, as accurately estimating sparse, high-power IN in real time is inherently difficult when only a limited number of null tones is available.
% if the estimated impulse is imperfectly subtracted, the resulting signal modification destroys the required orthogonality between subcarriers, directly generating self-induced ICI. 

Recent work has considered frequency-domain mitigation as an alternative approach. In \cite{Berka26}, a GMM is fitted to frequency-domain IN using a conventional EM algorithm, while the number of mixture components is selected using the Bayesian information criterion. The variance of the most likely component is then assigned to each sample and used as a proxy label for training a memory-aware neural network. The resulting diagonal noise-covariance estimates are incorporated into equalization and likelihood computation, thereby avoiding explicit reconstruction and subtraction of time-domain IN as in \cite{Caire08, Lampe11, Liu16, Lin11, Lin13, Korki16}.

However, the frequency-domain GMM in \cite{Berka26} is learned empirically rather than derived from the underlying time-domain Markov--Gaussian process. Consequently, its GMM-based labels represent EM estimates rather than analytically derived ground-truth frequency-domain NSI. Moreover, although its neural network processes the subcarriers sequentially to learn frequency-domain dependencies, the EM stage computes the posterior state probabilities independently for individual residual samples. It therefore neither assigns a common latent state to the null-tone observations within an OFDM symbol nor models state transitions between consecutive OFDM symbols. In contrast, this paper derives the scalar frequency-domain GMM induced by OFDM block processing, including its number of components, component variances, and mixing coefficients. Based on this characterization, we develop block-EM estimators that aggregate the null-tone observations within each OFDM symbol under a common latent state, while the HMM-EM variant additionally estimates the state-transition probabilities across consecutive OFDM symbols. The resulting NSI estimates are incorporated into state-dependent likelihood functions for robust frequency-domain detection. Our primary contributions are summarized as follows:
\begin{itemize}
    \item \emph{Statistical Characterization and Optimal Detection}: We derive a statistical representation of frequency-domain IN using a Gaussian mixture model (GMM), where its mixing coefficients, variances, and number of components are derived based on the block-processing nature of OFDM systems. By leveraging this model, we determine the conditions under which the central limit theorem (CLT) allows for a simplified AWGN receiver, and conversely, where non-Gaussianity necessitates precise NSI. On this basis, we derive the optimal frequency-domain receiver under perfect NSI to establish an upper bound for achievable performance.
    \item \emph{Block-EM Estimation Framework}: To bridge the gap between theory and practice, we propose a suite of EM algorithms that exploit the block-correlated behavior of frequency-domain IN. This includes an efficient symbol-by-symbol estimator and an enhanced HMM-based EM algorithm with learnable transition probabilities to capture high temporal correlation. 
    \item \emph{Adaptive Complexity via MAP-EM}: To address the state-space explosion inherent in large DFT sizes, we introduce a MAP-EM variant. By incorporating a sparsity-promoting prior on the mixing coefficients, this algorithm dynamically adapts the number of estimated noise states based on the observation window, ensuring robust NSI estimation without the overhead of a fixed, high-dimensional state space.
    \item \emph{System-Level Analysis and Trade-off Identification}: We identify the optimal use-case scenarios for each EM variant based on specific IN characteristics and system settings. Furthermore, we evaluate the critical trade-off between algorithmic complexity and the system overhead (i.e., the number of null tones), demonstrating where simplified models suffice and where sophisticated variants are necessary, particularly in resource-constrained settings.
\end{itemize}

The paper is organized as follows: Sec.~\ref{sec:sys} presents the transmission setup for our OFDM system. Sec.~\ref{sec:MMA} reviews the Markov--Middleton IN channel and its HMM-GMM representation in both time and frequency domains. In Sec.~\ref{sec:conv_rx} we illustrate the conventional coded-OFDM receiver design, while in Sec.~\ref{sec:em} we detail the three proposed block-EM algorithm designs for the NSI estimation. Sec.~\ref{sec:simu} presents the simulation results for the optimal receiver and the proposed block-EM-based receiver performance, and finally, Sec.~\ref{sec:conc} concludes this paper. 

\emph{Notation:} The notation convention used in this paper is defined as follows. Time-domain quantities are written in lowercase, while frequency-domain quantities are written in uppercase. Scalars denote individual entries (e.g., $x_k$ and $X_m$). Vectors are denoted by bold italic letters. For example, a time-domain vector from index $1$ to $K$ is written as $\boldsymbol{x}_1^K=[x_1, x_2, \ldots, x_K]^\mathsf{T}$, and vectors extracted from the $m$-th row and $n$-th column of a matrix are denoted by $\boldsymbol{X}_m$ and $\boldsymbol{X}^{(n)}$. Matrices are denoted by bold letters (e.g., $\mathbf{x}$ and $\mathbf{X}$), with $X_{m,n}$ representing the $(m,n)$-th entry. To reduce the notation overhead, we use $p(\cdot)$ to denote both the probability mass function for discrete variables and the PDF for continuous variables throughout this paper. A complex Gaussian distribution with mean $\mu$ and variance $\sigma^2$ is denoted as $\mathcal{CN}(\mu,\sigma^2)$, and its PDF evaluated at $x_t$ is written as $\mathcal{CN}(x_t;\mu,\sigma^2)$.

\section{OFDM Transmission System}\label{sec:sys}
\begin{figure*}
    % \vspace{-10mm}
    \centering
    \resizebox{1\textwidth}{!}{\includegraphics{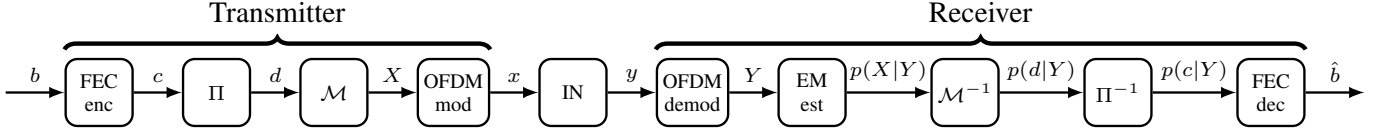}}
    \caption{System block diagram of a coded bit-interleaved PSK mapper with OFDM modulation on the transmitter side. The conventional receiver includes an OFDM demodulator, an EM estimator, and a feed-forward FEC decoding module.}
    \label{fig:sys_block}
\end{figure*}

%---------------------------
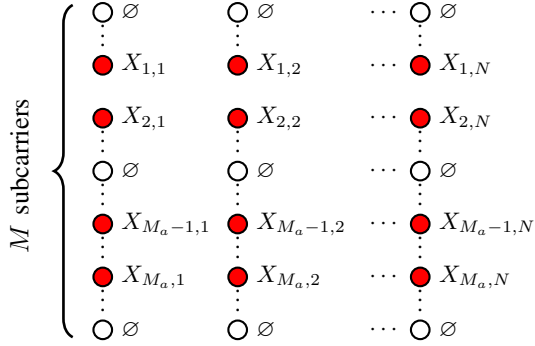
\begin{figure}[t]
    \centering
    {\begin{tikzpicture}[>=stealth]
\def\dx{2.1}   % horizontal step
\def\dy{0.7}   % vertical step
\tikzset{
  nodebase/.style={circle,draw,thick,inner sep=0pt,minimum size=7pt},
  zero/.style={nodebase,fill=white},
  xnode/.style={nodebase,fill=red,inner sep=1pt},
  xlab/.style={anchor=west,scale=0.9}
}

% --- 
\node[zero] (n-1-1) at (0*\dx,0*\dy) {};
\node[zero] (n-1-2) at (0.85*\dx,0*\dy) {};
\node[zero] (n-1-3) at (2*\dx,0*\dy) {};
\node[xlab, anchor = east] at (n-1-3.west) {$\cdots$};
\node[xlab] at (n-1-1.east) {$\varnothing$};
\node[xlab] at (n-1-2.east) {$\varnothing$};
\node[xlab] at (n-1-3.east) {$\varnothing$};

% --- 
\node[xnode] (n-2-1) at (0*\dx,-1*\dy) {};
\node[xnode] (n-2-2) at (.85*\dx,-1*\dy) {};
\node[xnode] (n-2-3) at (2*\dx,-1*\dy) {};
\node[xlab,anchor=south] at (n-2-1.north) {$\vdots$};
\node[xlab,anchor=south] at (n-2-2.north) {$\vdots$};
\node[xlab,anchor=south] at (n-2-3.north) {$\vdots$};
\node[xlab] at (n-2-1.east) {$X_{1,1}$};
\node[xlab] at (n-2-2.east) {$X_{1,2}$};
\node[xlab] at (n-2-3.east) {$X_{1,N}$};
\node[xlab, anchor = east] at (n-2-3.west) {$\cdots$};

% --- 
\node[xnode] (n-3-1) at (0*\dx,-2*\dy) {};
\node[xnode] (n-3-2) at (.85*\dx,-2*\dy) {};
\node[xnode] (n-3-3) at (2*\dx,-2*\dy) {};
\node[xlab] at (n-3-1.east) {$X_{2,1}$};
\node[xlab] at (n-3-2.east) {$X_{2,2}$};
\node[xlab, anchor = east] at (n-3-3.west) {$\cdots$};
\node[xlab] at (n-3-3.east) {$X_{2,N}$};

% --- 
\node[zero] (n-4-1) at (0*\dx,-3*\dy) {};
\node[zero] (n-4-2) at (.85*\dx,-3*\dy) {};
\node[zero] (n-4-3) at (2*\dx,-3*\dy) {};
\node[xlab,anchor=south] at (n-4-1.north) {$\vdots$};
\node[xlab,anchor=south] at (n-4-2.north) {$\vdots$};
\node[xlab,anchor=south] at (n-4-3.north) {$\vdots$};
\node[xlab, anchor = east] at (n-4-3.west) {$\cdots$};
\node[xlab] at (n-4-1.east) {$\varnothing$};
\node[xlab] at (n-4-2.east) {$\varnothing$};
\node[xlab] at (n-4-3.east) {$\varnothing$};

% --- 
\node[xnode] (n-5-1) at (0*\dx,-4*\dy) {};
\node[xnode] (n-5-2) at (.85*\dx,-4*\dy) {};
\node[xnode] (n-5-3) at (2*\dx,-4*\dy) {};
\node[xlab, anchor = east] at (n-5-3.west) {$\cdots$};
\node[xlab,anchor=south] at (n-5-1.north) {$\vdots$};
\node[xlab,anchor=south] at (n-5-2.north) {$\vdots$};
\node[xlab,anchor=south] at (n-5-3.north) {$\vdots$};
\node[xlab] at (n-5-1.east) {$X_{M_a-1,1}$};
\node[xlab] at (n-5-2.east) {$X_{M_a-1,2}$};
\node[xlab] at (n-5-3.east) {$X_{M_a-1,N}$};

% --- 
\node[xnode] (n-6-1) at (0*\dx,-5*\dy) {};
\node[xnode] (n-6-2) at (.85*\dx,-5*\dy) {};
\node[xnode] (n-6-3) at (2*\dx,-5*\dy) {};
\node[xlab,anchor=south] at (n-6-1.north) {$\vdots$};
\node[xlab] at (n-6-1.east) {$X_{M_a,1}$};
\node[xlab,anchor=south] at (n-6-2.north) {$\vdots$};
\node[xlab,anchor=south] at (n-6-3.north) {$\vdots$};
\node[xlab, anchor = east] at (n-6-3.west) {$\cdots$};
\node[xlab] at (n-6-2.east) {$X_{M_a,2}$};
\node[xlab] at (n-6-3.east) {$X_{M_a,N}$};
% --- 
\node[zero] (n-7-1) at (0*\dx,-6*\dy) {};
\node[zero] (n-7-2) at (.85*\dx,-6*\dy) {};
\node[zero] (n-7-3) at (2*\dx,-6*\dy) {};
\node[xlab,anchor=south] at (n-7-1.north) {$\vdots$};
\node[xlab,anchor=south] at (n-7-2.north) {$\vdots$};
\node[xlab,anchor=south] at (n-7-3.north) {$\vdots$};
\node[xlab, anchor = east] at (n-7-3.west) {$\cdots$};
\node[xlab] at (n-7-1.east) {$\varnothing$};
\node[xlab] at (n-7-2.east) {$\varnothing$};
\node[xlab] at (n-7-3.east) {$\varnothing$};
%---
\draw[line width=1pt, decorate,decoration={brace,amplitude=7pt, mirror}]
  ($(n-1-1.north west)+(-3mm,0)$) -- ($(n-7-1.south west)+(-3mm,0)$)
  node[pos=0.25,left=20pt, rotate = 90] {\normalsize $M$ subcarriers};

\end{tikzpicture}} 
    \caption{Transmitted symbol organization before IDFT. $\varnothing$ denotes the null-tone placement.}   
    \label{fig:2DX}
\end{figure}

As depicted in Fig.~\ref{fig:sys_block}, the information bits $\boldsymbol{b}_1^{KR}$ are processed by a rate-$R$ forward error correction (FEC) encoder and a bit-level interleaver ($\Pi$) to produce the interleaved coded sequence $\boldsymbol{d}_1^{K}$. A $C$-PSK mapper then converts these bits into a length-$T$ sequence $\boldsymbol{X}_1^T$, where
\begin{align}\label{eq:X}
    X_t \in \mathcal{X} = \{e^{j2\pi i/C} \mid i=0,1,\dots,C-1\}.
\end{align}

For OFDM modulation, $\boldsymbol{X}_1^T$ is structured into a matrix $\mathbf{X} \in \mathbb{C}^{M_a \times N}$ spanning $M_a$ active subcarriers and $N$ symbols. To suppress out-of-band emissions and DC leakage, $M_w$ guard null tones are inserted, yielding an augmented matrix of $M = M_a + M_w$ subcarriers (Fig.~\ref{fig:2DX}). An $M$-point inverse DFT generates the time-domain matrix $\mathbf{x} = \mathbf{F}^\mathsf{H} \mathbf{X}$. After prepending a length-$M_{cp}$ cyclic prefix (CP) to each symbol, $\mathbf{x}$ is serialized into the transmission vector $\boldsymbol{x}_1^{N(M+M_{cp})}$.

The system operates over a Markov--Middleton impulsive noise (IN) channel:
\begin{align}\label{eq:y_td}
    y_t = x_t + w_t,
\end{align}
where $w_t$ is the compound IN realization (detailed in Sec.~\ref{sec:td_mma}). At the receiver, following CP removal and reshaping into $\mathbf{y}\in\mathbb{C}^{M\times N}$, a DFT yields the frequency-domain signal $\mathbf{Y}=\mathbf{F}\mathbf{y}$, expressed element-wise as
\begin{align}\label{eq:Y_fd}
    Y_{m,n} = X_{m,n} + W_{m,n}.
\end{align}
The statistical properties of the frequency-domain noise $\mathbf{W}=\mathbf{F}\mathbf{w}$ are discussed in Sec.~\ref{sec:fd_mma}.

\section{Markov-Middleton Impulsive Noise Model}\label{sec:MMA}
The Markov--Middleton model~\cite{MMA} extends the memoryless Middleton Class A model~\cite{Middleton97} via an HMM process to capture the bursty IN observed in field measurements \cite{chc24_1, Sacuto14, MMA, Shan09}. In this section, we first outline the time-domain statistics of this model. We then derive the corresponding frequency-domain representations at the OFDM demodulator output, which is crucial for subsequent statistical analysis and receiver design.

\subsection{Time-Domain Representation}\label{sec:td_mma}
We first define the time-domain hidden noise state
\begin{align}\label{eq:state_td}
    s_t \in \mathcal{S}_t=\{0,1,\ldots,L-1\},
\end{align}
where $s_t=0$ denotes the background-noise state and $s_t>0$ denotes impulsive-noise states. The prior probability of state $j$ is
\begin{align}
    \pi_{t_j}=p(s_t=j), \nonumber
\end{align}
which is obtained by normalizing a truncated Poisson distribution with impulsive index $A$:
\begin{align}\label{eq:pti}
    \pi_{t_j}
    =
    \frac{e^{-A}A^j/j!}{\sum_{i=0}^{L-1} e^{-A}A^i/i!}.
\end{align}
Thus, the background state occurs with probability $\pi_B\approx e^{-A}$, while the impulsive states occur with total probability $\pi_I\approx 1-e^{-A}$. A larger $A$ therefore corresponds to more frequent impulsive events.

Conditioned on the state $s_t=j$, the time-domain noise sample is a single Gaussian:
\begin{align}
    p(w_t\mid s_t=j)
    =
    \mathcal{CN}(w_t; 0,\sigma^2_{t_j}), \nonumber
\end{align}
where the state-dependent variance is defined as
\begin{align}\label{eq:varm}
    \sigma_{t_j}^2
    &=
    \sigma_B^2+\frac{j}{A}\sigma_I^2
    =
    \left(1+\frac{j\Lambda}{A}\right)\sigma_B^2,
\end{align}
where $\sigma_B^2\triangleq\sigma_{t_0}^2$ is the background Gaussian
variance, $\sigma_I^2$ is the mean impulsive-component power of the
underlying Middleton Class-A model, and
$\Lambda\triangleq\sigma_I^2/\sigma_B^2$ is the corresponding
impulsive-to-background power ratio. Thus, the conditional variance in
state $j$ consists of the background variance and the excess impulsive
variance $(j/A)\sigma_I^2$.
% where the state-dependent variance is defined as
% \begin{align}\label{eq:varm}
%     \sigma^2_{t_j}
%     =
%     \sigma^2_{t_0}+(j\Lambda/A)\sigma^2_{t_0}
% \end{align}
% Here, $\Lambda=\sigma_I^2/\sigma_B^2$ is the user-defined impulsive-to-background average power ratio, with $\sigma_B^2=\pi_{t_0}\sigma^2_{t_0}$ and $\sigma_I^2=\sum_{j=1}^{L-1}\pi_{t_j}\sigma^2_{t_j}$. In \eqref{eq:varm}, the background noise power remains present in every state, while $\Lambda$ scales the state-dependent excess impulsive variance. 
For a fixed $\Lambda$, decreasing $A$ reduces the occurrence of impulsive events but increases their individual variances, as illustrated in Figs.~\ref{fig:noise_a01r0} and \ref{fig:noise_a005r0}.

Marginalizing over the hidden state gives the time-domain GMM:
\begin{align}\label{eq:pn}
    p(w_t)
    =
    \sum_{j=0}^{L-1}
    \pi_{t_j}\mathcal{CN}(w_t; 0,\sigma^2_{t_j}).
\end{align}
Therefore, the time-domain state $s_t$ identifies which Gaussian component, and hence which noise variance, generates the sample $w_t$.

To model burstiness, the hidden state sequence follows a first-order Markov chain with transition probabilities~\cite{MMA}
\begin{align}\label{eq:P}
    P_{t_{ij}}
    \triangleq
    p(s_t=j\mid s_{t-1}=i)
    =
    \begin{cases}
        r+(1-r)\pi_{t_j}, & i=j,\\
        (1-r)\pi_{t_j}, & i\neq j,
    \end{cases}
\end{align}
where $r\in[0,1]$ is the correlation parameter. 
A larger $r$ increases the probability of remaining in the same state, producing highly correlated noise samples (e.g., $r=0.9$ in Fig.~\ref{fig:noise_a005r09}). Conversely, setting $r=0$ reverts to the memoryless Middleton Class A model (Figs.~\ref{fig:noise_a01r0}--\ref{fig:noise_a005r0}). In the limiting case $r=1$, the realizations exhibit a joint GMM behavior, where each block of samples is drawn from a single Gaussian component of the mixture. 

In summary, the Markov--Middleton model is fully defined by four parameters: the background noise variance $\sigma^2_{t_0}$, the impulsive index $A$, the power ratio $\Lambda$, and the correlation parameter $r$.

\iffigureson
\begin{figure*}[t]
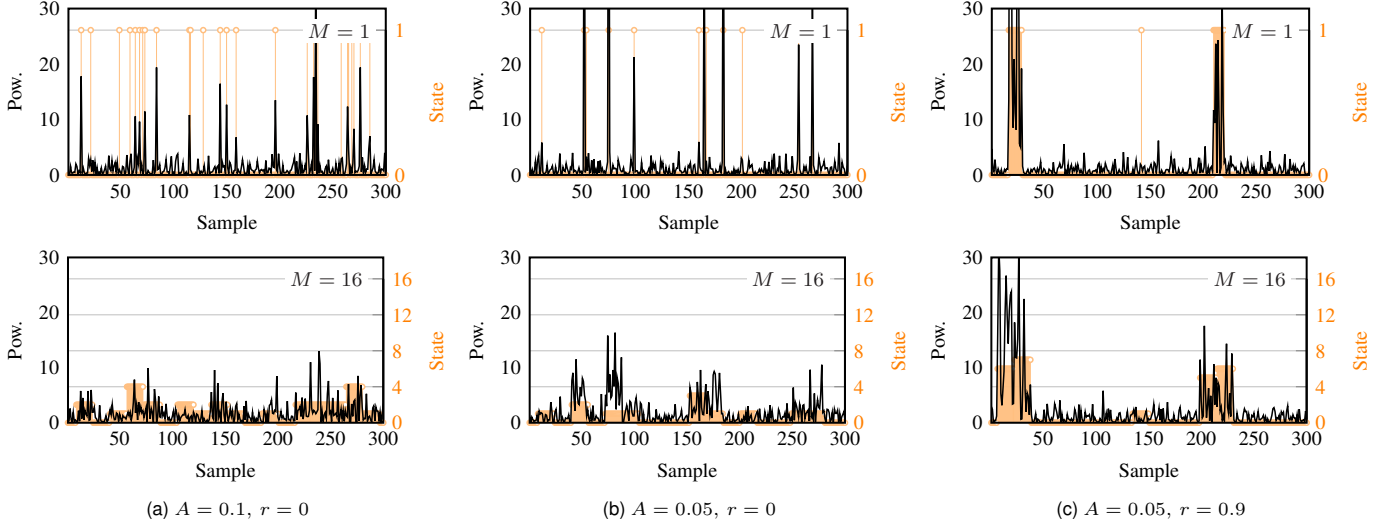

\centering
    \subfloat[$A=0.1,\, r=0$ \label{fig:noise_a01r0}]{%
      \begin{minipage}{.33\textwidth}
        \centering
        \resizebox{1\columnwidth}{!}{\includegraphics{tikz/noise_r0N1.tikz}}\\
        % \resizebox{1\columnwidth}{!}{\includegraphics{tikz/noise_r0N4.tikz}}\\
        \resizebox{1\columnwidth}{!}{\includegraphics{tikz/noise_r0N16.tikz}}
      \end{minipage}
    }
    \subfloat[$A=0.05,\, r=0$ \label{fig:noise_a005r0}]{%
      \begin{minipage}{.33\textwidth}
        \centering
        \resizebox{1\columnwidth}{!}{\includegraphics{tikz/noise_r0A005N1.tikz}}\\
        % \resizebox{1\columnwidth}{!}{\includegraphics{tikz/noise_r0A005N4.tikz}}\\
        \resizebox{1\columnwidth}{!}{\includegraphics{tikz/noise_r0A005N16.tikz}}
      \end{minipage}
    }
    \subfloat[$A=0.05,\, r=0.9$ \label{fig:noise_a005r09}]{%
      \begin{minipage}{.33\textwidth}
        \centering
        \resizebox{1\columnwidth}{!}{\includegraphics{tikz/noise_r09N1.tikz}}\\
        % \resizebox{1\columnwidth}{!}{\includegraphics{tikz/noise_r09N4.tikz}}\\
        \resizebox{1\columnwidth}{!}{\includegraphics{tikz/noise_r09N16.tikz}}
      \end{minipage}
    }
    \label{fig:MMA_noise}
    \caption{Noise realization from a two-state Markov--Middleton impulsive noise model with a fixed $\sigma^2_{t_0}=1$ and $\Lambda=1$ for block sizes of $M=1$ and $M=16$ with (a) $A=0.1, r=0$, (b) $A=0.05, r=0$, and (c) $A=0.05, r=0.9$. The orange vertical bar denotes the index $j$ of the state realizations \( s_t = j \) for generating the corresponding noise samples.}
\end{figure*}
\fi

% FD representation======================================================================
\subsection{Frequency-Domain Representation}\label{sec:fd_mma}
In this subsection, we focus on the statistical representation of the frequency-domain IN $W_{m,n}$ in \eqref{eq:Y_fd}. 

For the $n$-th OFDM symbol, let $\boldsymbol{s}^{(n)}=(s_{0,n},s_{1,n},\ldots,s_{M-1,n})$ denote the time-domain state sequence over the $M$ samples entering the DFT. 
Since the DFT is linear, the scalar frequency-domain noise sample is
\begin{align}
    W_{m,n}=\sum_{\ell=0}^{M-1}F_{m,\ell}w_{\ell,n}, \nonumber
\end{align}
where 
$$ p(w_{\ell,n} \mid s_{\ell,n}=j) = \mathcal{CN}(w_{\ell,n}; 0, \sigma^2_{t_j}). $$

Using the normalized DFT matrix, $|F_{m,\ell}|^2=1/M$, the conditional variance of $W_{m,n}$ can be written as a count-weighted average of the time-domain state variances:
\begin{align} \label{eq:pWs}
    p(W_{m,n}\mid \boldsymbol{s}^{(n)})
    = \mathcal{CN}\left(W_{m,n}; 0, \sigma^2_{f_j} \right), 
\end{align}
where
\begin{align}\label{eq:var_f}
    \sigma^2_{f_j} = \frac{1}{M}\sum_{i=0}^{L-1}c_i^j\sigma^2_{t_i}.
\end{align}

In \eqref{eq:var_f}, $c_i^j$ is the number of samples in the block whose time-domain state is $i$. The collection of these counts forms the state-count vector 
$$ \mathbf{c}^j = (c^j_0, c^j_1, \dots, c^j_{L-1}) \in \mathcal{C}_M, \quad \sum_{i=0}^{L-1} c^j_i=M. $$ Consequently, for a scalar frequency-domain sample, the associated variance depends on the time-domain state sequence exclusively through the number of samples drawn from each time-domain state. This count-weighted frequency-domain variance formulation holds exactly for the Markov--Middleton HMM under conditionally independent Gaussian assumptions.

Accordingly, we define the frequency-domain noise state 
\begin{align}\label{eq:state_fd}
    S_{m,n} \in \mathcal{S}_f = \{0, 1, \ldots, L_f-1\}
\end{align}
such that $S_{m,n}=j$ corresponds to the case where the underlying time-domain sequence $\boldsymbol{s}^{(n)}$ induces the $j$-th count vector $\mathbf{c}^j$. Thus, \eqref{eq:pWs} can be equivalently expressed as
\begin{align}
    p(W_{m,n}\mid S_{m,n}=j)
    = \mathcal{CN}\left(W_{m,n}; 0, \sigma^2_{f_j} \right). \nonumber
\end{align}

Consequently, the marginal scalar distribution of the frequency-domain IN is the transformed GMM
\begin{align}\label{eq:fN}
    p(W_{m,n})
    =
    \sum_{j=0}^{L_f-1}
    \pi_{f_j}\mathcal{CN}(W_{m,n}; 0, \sigma^2_{f_j}),
\end{align}
where the number of possible frequency-domain states $L_f$ equals the number of valid state-count vectors
\begin{align} \label{eq:Wf}
    L_f=|\mathcal{C}_M|=\binom{M+L-1}{L-1}.
\end{align}
This implies the frequency-domain dimension $L_f$ can be significantly larger than $L$. This cardinality is unaffected by the correlation parameter $r$; however, the probability assigned to each count vector depends on the temporal state model.

The mixing coefficients $\pi_{f_j}$ in \eqref{eq:fN} are the probabilities of the corresponding state-count vectors
\begin{align}\label{eq:Prif}
    \pi_{f_j}=p(S_{m,n}=j)=p(\mathbf{c}^j)
\end{align}

For independent time-domain states, i.e., $r=0$, this count probability reduces to the multinomial form
\begin{align}\label{eq:Prif_iid}
    \pi_{f_j}
    =
    \frac{M!}{c_0^j!c_1^j!\cdots c_{L-1}^j!}
    \prod_{i=0}^{L-1}\pi_{t_i}^{c_i^j}.
\end{align}

For Markov-correlated states, a frequency-domain count vector is no longer associated with a unique probability through a multinomial distribution. The reason is that the same count vector can be generated by multiple ordered state sequences, and these sequences generally have different probabilities because of the Markov memory. For example, two sequences may contain the same number of background and impulsive states but differ in the number of state transitions. Since the transition probabilities depend on the correlation parameter $r$, their occurrence probabilities are generally not equal.

Assuming the Markov chain is in steady state, the probability of one ordered time-domain state sequence 
$$\boldsymbol{s}=(s_0,s_1,\ldots,s_{M-1})$$ 
is
\begin{align}\label{eq:path_prob}
    p(\boldsymbol{s})
    =
    \pi_{t_{s_0}}
    \prod_{\ell=1}^{M-1}
    p(s_\ell\mid s_{\ell-1}),
\end{align}
where $p(s_\ell\mid s_{\ell-1})$ is given by \eqref{eq:P}. 

Let $Q(\mathbf{c}^j)$ denote the set of all ordered state sequences whose state-count vector equals $\mathbf{c}^j$. Since every sequence in $Q(\mathbf{c}^j)$ produces the same frequency-domain variance but may occur with a different probability, the probability of the count vector is obtained by summing the probabilities of all such sequences as
\begin{align}\label{eq:Prif_markov}
    \pi_{f_j}
    =
    \sum_{\boldsymbol{s}\in\mathcal{Q}(\mathbf{c}^j)}
    p(\boldsymbol{s}).
\end{align}

Consequently, the multinomial expression \eqref{eq:Prif_iid} is exact only for the memoryless case ($r=0$). For the Markov--Middleton case ($r>0$), the count-vector probabilities depend on the transition matrix through \eqref{eq:path_prob}--\eqref{eq:Prif_markov}. Importantly, the set of possible count vectors, and therefore the frequency-domain state cardinality and the associated variances remain unchanged. Only their probabilities are modified by the temporal correlation

While practical IN channels may exhibit numerous states, assuming $L=2$ is sufficient to capture the channel behavior for robust receiver design, incurring only marginal performance degradation \cite{MMA, chc25_1}. Under this two-state assumption, the frequency-domain state space becomes $L_f=M+1$. In the independent case, the mixing coefficient in \eqref{eq:Prif_iid} simplifies to the binomial distribution:
\begin{align}\label{eq:Prif_bi}
    \pi_{f_j} = \binom{M}{c^j_0} \pi_{t_0}^{c^j_0} \pi_{t_1}^{M-c^j_0},
\end{align}
whereas the Markov-correlated case has the same $M+1$ count states but uses the Markov count probability in \eqref{eq:Prif_markov}.

Fig.~\ref{fig:pdf} depicts the PDF evolution for this two-state model under independent time-domain noise assumptions. At $M=1$ (the time-domain case), the background and impulsive states are highly distinct. As $M$ increases (e.g., to $M=16$), the model expands to $L_f=17$ components. Despite this expansion, the probability mass remains highly concentrated; for instance, the first five components alone account for over $90\%$ of the noise realizations.

To capture inter-block noise memory, the transition probability from state-count vector $\mathbf{c}^i$ to $\mathbf{c}^j$ in the subsequent block is defined as
\begin{align}\label{eq:P_block}
    P^b_{f_{ij}} \triangleq p(\mathbf{c}^j \mid \mathbf{c}^i).
\end{align}
Deriving $P^b_{f_{ij}}$ analytically requires marginalizing over all $M$-step paths through the time-domain Markov chain to account for sample-by-sample transitions across block boundaries. Due to the rapid expansion of the state space $\mathcal{C}_M$ and the combinatorial complexity of evaluating these conditional paths, an exact computation of \eqref{eq:P_block} is intractable for practical block sizes. Therefore, in Sec.~\ref{sec:em_hmm}, a block-EM-based algorithm is introduced to estimate this block-wise transition matrix directly from received data.

%%============================================================================
\iffigureson
\begin{figure}
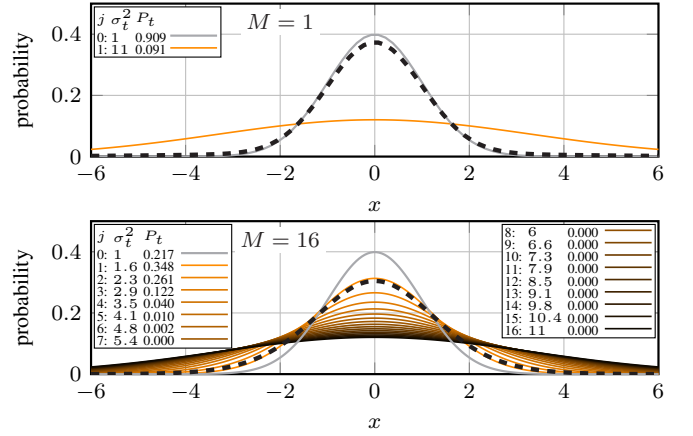

    \centering
    \begin{minipage}{1\columnwidth}
        \centering
        \resizebox{1\columnwidth}{!}{\includegraphics{tikz/pdf_n1.tikz}}\\
        \resizebox{1\columnwidth}{!}{\includegraphics{tikz/pdf_n16.tikz}}
    \end{minipage}
    \caption{Normalized PDFs of frequency-domain impulsive noise for $M=1$ and $M=16$ under a 2-state Markov--Middleton noise model with parameters $A=0.1$ and $\Lambda=1$. The dashed black curve represents the total PDF of all the states weighted by their mixing coefficients.}
    \label{fig:pdf}
\end{figure}
\fi
%%============================================================================

\subsection{Noise State Information and Scalar Likelihood Representation}
\label{sec:nsi_likelihood}
The hidden states defined in \eqref{eq:state_td} and \eqref{eq:state_fd} constitute the NSI used by the receiver. In the time domain, $s_t$ identifies the Gaussian component that generates the noise sample $w_t$, thereby determining the corresponding variance $\sigma^2_{t_j}$. 

In the frequency domain, the state is no longer associated with an individual sample. Instead, it is induced by the complete DFT block. Specifically, the ordered time-domain state sequence $\boldsymbol{s}^{(n)}$ associated with the $n$-th OFDM symbol produces a state-count vector $\mathbf{c}^j$, which in turn determines the frequency-domain variance $\sigma^2_{f_j}$ in \eqref{eq:var_f}. The frequency-domain state therefore represents the aggregate composition of time-domain noise states within an OFDM symbol, rather than an independently generated hidden state on each subcarrier.

This observation motivates a block-level state representation. Conditioned on the state associated with count vector $\mathbf{c}^j$, every subcarrier in the same OFDM symbol has a scalar marginal Gaussian distribution with identical variance:
\begin{align} \label{eq:fN_scalar}
    p(W_{m,n}\mid S_{n}=j)
    =
    \mathcal{CN}(W_{m,n}; 0,\sigma^2_{f_j}).
\end{align}
Although the likelihood is evaluated separately for each subcarrier observation $W_{m,n}$, the latent state itself is common to the entire OFDM symbol and is therefore represented by the block-level state $S_n$. Consequently, all subcarriers within the same OFDM symbol share the same state-dependent variance parameter $\sigma_{f_j}^{2}$, while their individual noise realizations remain different.

Because the count vector affects receiver processing only through the resulting variance ($\sigma_{f_j}^{2}$), the block-level state retains the relevant NSI information. This representation supports NSI estimation from null-tone observations while yielding a tractable likelihood and facilitating the practical frequency-domain receiver design that will be further discussed in Sec.~\ref{sec:em}.

% This paper focus on frequency-domain receiver design, and the scalar likelihood representation in \eqref{eq:fN_scalar} is sufficient for this purpose. The block-level state $S_n$ captures the dominant state-dependent power information needed for NSI estimation from null tones, while enabling a tractable likelihood and a straightforward frequency-domain receiver design that will be further discussed in Sec.~\ref{sec:em}.

\section{Conventional OFDM Receiver}\label{sec:conv_rx}
Following OFDM demodulation, the received frequency-domain signal $Y_{m,n}$ from $\eqref{eq:Y_fd}$ is used to compute the observation likelihood function. In conventional receivers, this function is calculated under the simplifying assumption of a single AWGN component for all received samples:
\begin{align}\label{eq:lik_con}
     p(Y_{m,n} \mid X_{m,n}=\mu_i) = \mathcal{CN}(Y_{m,n}; \mu_i, \hat{\sigma}_f^2),
\end{align}
where $\mu_i$ is drawn from the modulated symbol space defined in \eqref{eq:X}, and $\hat{\sigma}_f^2$ is the estimated noise variance, typically obtained using a simple power-based technique over the null-tones. Given the prior symbol probability $p(X_{m,n})$, the a posteriori probability (APP) of the transmitted symbol conditioned on the observation $Y_{m,n}$ is obtained via Bayes' theorem 
\begin{align}\label{eq:sym_pos_con}
    p(&X_{m,n}=\mu_i \mid Y_{m,n})  \nonumber \\
    &=\frac{p(X_{m,n}=\mu_i) p(Y_{m,n} \mid X_{m,n}=\mu_i)}{\sum_{j}p(X_{m,n}=\mu_{j}) p(Y_{m,n} \mid X_{m,n}=\mu_{j})}.
\end{align}
Subsequently, the symbol APP is converted into the bit APP for a specific coded bit $d_k$ through the symbol-to-bit demapping process ($\mathcal{M}^{-1}$). This conversion is achieved by marginalizing the symbol APPs over all symbols $\mu_i$ that share a specific value $j$ for that bit as
\begin{align}
    p(d_k=j \mid Y_{m,n}) = \sum_{\mu_i \in \mathcal{X}_j^{k}} p(X_{m,n}=\mu_i \mid Y_{m,n} ), \quad j \in \{0, 1\} \nonumber
\end{align}
where $\mathcal{X}_j^{k}$ is the set of all symbols $\mu_i$ in the constellation that have the value $j$ for their $k$-th bit. 

A bit-level deinterleaver ($\Pi^{-1}$) is then applied to reverse the permutation imposed during transmission, recovering the posterior coded bit probability:
\begin{align}\label{eq:c_Y}
    p(c_k\mid Y_{m,n}) = \Pi^{-1}\left[p(d_k \mid Y_{m,n})\right].    
\end{align}
Finally, the entire sequence of APPs 
$\{p(c_k \mid Y_{m,n})\}_{k=1}^K$
is fed to the FEC decoder to estimate the information bit sequence:
\begin{align}  \label{eq:jointc}
    \hat{\boldsymbol{b}}_1^{KR}= \mathcal{F}\left[\{p(c_k\mid Y_{m,n})\}_{k=1}^K\right]
\end{align}
where $\mathcal{F}$ represents the full FEC decoding function that processes the APP sequence to produce the final estimate of the information bits.

\section{Block-EM-based NSI Estimation}\label{sec:em}
In Sec.~\ref{sec:conv_rx}, we introduced the conventional OFDM receiver. However, its reliance on the AWGN assumption in the likelihood function \eqref{eq:lik_con} is fundamentally suboptimal when the channel exhibits impulsive noise. 

In contrast, the optimal likelihood function requires perfect knowledge of the frequency-domain NSI, using the true noise state $S_{n}=j$ to select the correct variance as
\begin{align} \label{eq:lik_per}
    p(Y_{m,n} \mid X_{m,n}=\mu_i, S_{n}=j) = \mathcal{CN}(Y_{m,n}; \mu_i, {\sigma}_{f_j}^2).
\end{align}
This conditional likelihood represents the ideal scenario where the receiver knows exactly which state $S_{n}$ determines the noise power $\sigma_{f_j}^2$. 

In practice, the NSI is hidden and is not directly observable, making its inference from the received data the central challenge. Accordingly, this paper investigates EM-based estimation techniques for NSI using null tones in the OFDM system, which serves as the primary focus of the subsequent sections.

\subsection{Symbol-by-symbol block-EM: SBS-EM}\label{sec:em_std}
The conventional EM algorithm is a well-established framework for maximum likelihood estimation in the presence of latent variables (see \cite{HMM}, \cite[Ch.~9]{Bishop}, and references therein). Its core mechanism involves iteratively alternating between an E-step, which computes the posterior distribution of the hidden states, and an M-step, which updates the model parameters to maximize the expected log-likelihood.

Motivated by the block-behavior of frequency-domain IN analyzed in Sec.~\ref{sec:fd_mma}, a block-EM framework is developed. Unlike conventional approaches that treat noise samples independently, the proposed method leverages the collective information across all samples within an OFDM symbol to compute an OFDM-symbol-based likelihood, assigning a single state to characterize the entire block. The discrete state space for the proposed block-based EM algorithm is defined as
\begin{align}\label{eq:state_em}
    \hat{S}_{n} \in \hat{\mathcal{S}}= \{0,1,\cdots,L_{\text{EM}}-1\},   
\end{align}
where $\hat{S}_{n}$ denotes the state representing the entire $n$-th OFDM symbol and $L_{\text{EM}}$ serves as a tunable design parameter that facilitates a flexible trade-off between estimation resolution and computational complexity. By employing this constrained state space, the algorithm provides a numerically tractable approximation of the true, high-dimensional frequency-domain NSI states defined in \eqref{eq:state_fd}.

For the SBS-EM, the target posterior distribution is defined as $p(\hat{S}_{n} \mid \boldsymbol{W}^{(n)})$, where $\boldsymbol{W}^{(n)}$ represents the null-tone observations for the $n$-th OFDM symbol. The parameter set estimated during the M-step is
\begin{align}\label{eq:hmmpara}
    \theta_{\text{SBS}} = \{\hat{\sigma}^2_{f_j}, \hat{\pi}_{f_{j}}\}, \quad j \in \{0, 1, \dots, L_{\text{EM}}-1\},
\end{align}
where $\hat{\sigma}^2_{f_j}$ and $\hat{\pi}_{f_{j}}$ denote the SBS-EM estimated frequency-domain variance and mixing coefficient of the GMM defined in \eqref{eq:fN}, respectively.

Following the scalar-likelihood discussed in Sec.~\ref{sec:nsi_likelihood}, the approximated block-likelihood function is written as
\begin{align}
    p(\boldsymbol{W}^{(n)} \mid \hat{S}_n=j) &\approx \prod_{m=1}^{M_w} p({W}_{m,n} \mid \hat{S}_n=j) \nonumber \\
    &= \prod_{m=1}^{M_w} \mathcal{CN}(W_{m,n}; 0,\hat{\sigma}^2_{f_j}). \label{eq:lik_em1}
\end{align}
The joint distribution follows as
\begin{align}
    p(\boldsymbol{W}^{(n)},\hat{S}_n=j) = \hat{\pi}_{f_j} \cdot p(\boldsymbol{W}^{(n)} \mid \hat{S}_n=j),
\end{align}
where $\hat{\pi}_{f_j}$ denotes the estimated prior state probability (i.e., GMM mixing coefficient). The symbol-by-symbol posterior probability is then expressed as 
\begin{align}\label{eq:pos_em1}
    p(\hat{S}_n=j \mid \boldsymbol{W}^{(n)}) = \frac{p(\boldsymbol{W}^{(n)},\hat{S}_n=j)}{\sum_{{\hat{S}_n} \in \hat{\mathcal{S}}} p(\boldsymbol{W}^{(n)},\hat{S}_n)}.  
\end{align}

Using the posterior probabilities derived in \eqref{eq:pos_em1}, the M-step updates the frequency-domain prior probability and the state-dependent variance for each state $j$ as follows:
\begin{align}
    \hat{\pi}_{f_j} &= \frac{1}{N} \sum_{n=1}^{N} p(\hat{S}_n = j \mid \boldsymbol{W}^{(n)}), \label{eq:em1_pri}  \\
    \hat{\sigma}^2_{f_j} &= \frac{\sum_{n=1}^N \left[ p(\hat{S}_n = j \mid \boldsymbol{W}^{(n)})  \frac{1}{M_w} \sum_{m=1}^{M_w} |W_{m,n}|^2 \right]}{\sum_{n=1}^N p(\hat{S}_n = j \mid \boldsymbol{W}^{(n)})}, \label{eq:em1_var}
\end{align}
respectively. These updated parameters, $\hat{\pi}_{f_j}$ and $\hat{\sigma}^2_{f_j}$, are then substituted back into \eqref{eq:lik_em1}--\eqref{eq:pos_em1} for the subsequent iteration, iteratively refining the posterior estimates until convergence.

% HMM-based block-EM: HMM-EM=================================================
\subsection{HMM-based block-EM: HMM-EM}\label{sec:em_hmm}
As illustrated in Fig.~\ref{fig:noise_a005r09}, when IN exhibits a high correlation coefficient $r$ or the OFDM system employs a relatively small DFT size $M$, the frequency-domain IN demonstrates significant temporal correlation across consecutive OFDM symbols. Under these conditions, the memoryless, symbol-by-symbol block-EM approach described in Sec.~\ref{sec:em_std} fails to accurately characterize the inter-OFDM symbol transition probabilities. To address this limitation, an HMM-based block-EM framework is derived in this section. This approach explicitly models the latent transition dynamics to enhance the accuracy of NSI tracking.

The latent states are now assumed to follow a first-order Markov process. This assumption enables the use of the BCJR algorithm \cite{HMM, bcjr74} to efficiently compute the joint distribution of the state at index $n$ and the full observation set of an OFDM frame, $\mathbf{W} \in \mathbb{C}^{M_w \times N}$. The joint probability is given by
\begin{align} 
     p( \hat{S}_n,  \mathbf{W}) &= \sum_{ \hat{S}_{n-1} \in \hat{\mathcal{S}}}{p(\hat{S}_{n}, \hat{S}_{n-1}, \mathbf{W})} \label{eq:map1} \\ 
    &= \sum_{ \hat{S}_{n-1} \in \hat{\mathcal{S}}}{\alpha(\hat{S}_{n-1})} \gamma(\hat{S}_n, \hat{S}_{n-1}) {\beta(\hat{S}_n)} \label{eq:map2},
\end{align}
where $\alpha(\hat{S}_{n})$ and $\beta(\hat{S}_n)$ denote the forward and backward recursions, respectively, and are defined as
\begin{align} 
    &\alpha(\hat{S}_{n}) = \sum_{\hat{S}_{n-1}\in\hat{\mathcal{S}}} \gamma(\hat{S}_{n}, \hat{S}_{n-1}) \cdot \alpha(\hat{S}_{n-1})  \\ 
    &\beta(\hat{S}_{n-1})= \sum_{\hat{S}_{n} \in \hat{\mathcal{S}}} \beta(\hat{S}_n) \cdot \gamma(\hat{S}_{n}, \hat{S}_{n-1}).
\end{align}
The branch metric $\gamma(\hat{S}_{n}, \hat{S}_{n-1})$ is expressed as
\begin{align}
    \gamma(\hat{S}_{n}, \hat{S}_{n-1}) &=  p(\boldsymbol{W}^{(n)}, \hat{S}_n \mid \hat{S}_{n-1}) \nonumber \\   
    & = p(\boldsymbol{W}^{(n)} \mid \hat{S}_n) \cdot \hat{P}^b_{f_{ij}}, \label{eq:gamma}
\end{align}
where the first term is the block likelihood function derived in \eqref{eq:lik_em1}, and the second term $\hat{P}^b_{f_{ij}} = p(\hat{S}_n \mid \hat{S}_{n-1})$ denotes the state transition probability between adjacent OFDM symbols. 
For a comprehensive derivation of the BCJR algorithm and its specific application to bursty impulsive noise environments, the reader is referred to \cite{HMM, bcjr74} and \cite{Dario09, MMA, chc25_1, chc25_2}, respectively.

After obtaining the joint probability \eqref{eq:map1} through \eqref{eq:map2}--\eqref{eq:gamma}, we can compute the posterior probability of state $\hat{S}_n$ given the entire null-tone observation within one OFDM frame $\mathbf{W}$ via
\begin{align} \label{eq:pos_em2}
    p(\hat{S}_n=j \mid \mathbf{W}) = \frac{p(\hat{S}_n=j, \mathbf{W})}{\sum_{\hat{S}_n \in \hat{\mathcal{S}}} p(\hat{S}_n,  \mathbf{W})}.
\end{align}
Compared to \eqref{eq:pos_em1}, the HMM-EM uses the entire observation space to compute the posterior state probability in \eqref{eq:pos_em2}, which is subsequently used to update the prior state probability in \eqref{eq:em1_pri} and the state-dependent variance in \eqref{eq:em1_var}. 

Furthermore, the state transition probability between adjacent OFDM symbols is learned via
\begin{align}
    \hat{P}^b_{f_{ij}} = \frac{ \sum_{n=1}^{N}{p(\hat{S}_n=j,\hat{S}_{n-1}=i \mid \mathbf{W})}} {\sum_{n=1}^N{\sum_{\hat{S}_n \in \hat{\mathcal{S}}}{p( \hat{S}_n, \hat{S}_{n-1}=i \mid \mathbf{W})} }}, \nonumber
\end{align}
where the posterior over the state pair conditioned on the entire observation space is given by
\begin{align}
    p(\hat{S}_n,\hat{S}_{n-1} \mid \mathbf{W}) = \frac{p(\hat{S}_n,\hat{S}_{n-1}, \mathbf{W}) }{\sum_{\hat{S}_n, \hat{S}_{n-1} \in \hat{\mathcal{S}}}p(\hat{S}_n, \hat{S}_{n-1}, \mathbf{W})}. \nonumber
\end{align}
Therefore, the learned parameter set in the HMM-EM framework is 
\begin{align}\label{eq:hmmpara_em2}
    \theta_{\text{HMM}} = \{\hat{\sigma}^2_{f_j}, \hat{\pi}_{f_{j}}, \hat{P}^b_{f_{ij}} \}, \quad j \in \{0, 1, \dots, L_{\text{EM}}-1\}.
\end{align}

%% MAP-based block-EM: MAP-EM=================================================
\subsection{MAP-based block-EM: MAP-EM}\label{sec:em_var}
Building on the fixed-state SBS-EM and HMM-EM frameworks in Secs.~\ref{sec:em_std} and \ref{sec:em_hmm}, the remaining design issue is how to choose a tractable estimated state space. As the DFT size increases, the analytical number of frequency-domain states grows rapidly, increasing computational complexity and making finite-state EM estimation less reliable. However, this growth does not imply that all analytical states should be recovered. Since the frequency-domain variance is a count-weighted average of time-domain variances in \eqref{eq:var_f}, different count states can produce closely spaced variance values, and this separation decreases as the DFT block size increases.

Therefore, the objective of MAP-EM is not to identify the exact analytical state space or a universally optimal number of states. Instead, it seeks a compact set of representative states whose variances preserve the dominant NSI needed for receiver detection while keeping the estimation problem manageable. To this end, a Dirichlet prior is imposed on the mixing coefficients to promote sparsity and prune weakly supported states during the EM iterations.

Instead of relying on the frequentist point estimate of the mixing coefficients as in \eqref{eq:em1_pri}, a Bayesian formulation is adopted by introducing a prior distribution. Specifically, the mixing coefficients $\boldsymbol{\hat{\pi}}_f = [\hat{\pi}_{f_0}, \dots, \hat{\pi}_{f_{L_{\text{EM}}-1}}]^T$ are modeled using a symmetric Dirichlet distribution:
\begin{align}
    p(\boldsymbol{\hat{\pi}}_f) = \text{Dir}(\boldsymbol{\hat{\pi}}_f \mid \eta) = \frac{1}{B(\eta)} \prod_{j=0}^{L_{\text{EM}}-1} \hat{\pi}_{f_j}^{\eta-1}, \nonumber
\end{align}
where $B(\eta)$ is the multivariate beta function serving as the normalization constant, and $\eta$ denotes the concentration parameter. By setting $\eta > 1$, the prior biases the estimate toward a uniform distribution, whereas choosing $0 < \eta < 1$ promotes sparsity, allowing the MAP-EM algorithm to effectively prune states that do not contribute meaningfully to the observed noise statistics.

The MAP estimate for the M-step update of the mixing coefficients is now derived. First, the likelihood of the latent states $\boldsymbol{\hat{S}}_1^N$, given the mixing coefficient vector $\boldsymbol{\hat{\pi}}_f$, can be expressed as a multinomial distribution:
\begin{align}
    p(\boldsymbol{\hat{S}}_1^N \mid \boldsymbol{\hat{\pi}}_f) &= \prod_{n=1}^N p(\hat{S}_n \mid \boldsymbol{\hat{\pi}}_f) \nonumber \\
    &= \prod_{n=1}^N \prod_{j=0}^{L_{\text{EM}}-1} \hat{\pi}_{f_j}^{\mathbb{I}(\hat{S}_n=j)} = \prod_{j=0}^{L_{\text{EM}}-1} \hat{\pi}_{f_j}^{N_j}, \label{eq:lik_map}
\end{align}
where $\mathbb{I}(\cdot)$ is the indicator function and 
$N_j = \sum_{n=1}^N p(\hat{S}_n = j \mid \boldsymbol{W}^{(n)})$ 
represents the effective number of observations (expected counts) assigned to state $j$ during the E-step.

Because the multinomial likelihood in \eqref{eq:lik_map} is conjugate to the Dirichlet prior, the posterior distribution $p(\boldsymbol{\hat{\pi}}_f \mid \boldsymbol{\hat{S}}_1^N)$ is also Dirichlet-distributed
\begin{align}
    p(\boldsymbol{\hat{\pi}}_f \mid \boldsymbol{\hat{S}}_1^N) &\propto p(\boldsymbol{\hat{S}}_1^N \mid \boldsymbol{\hat{\pi}}_f) \cdot p(\boldsymbol{\hat{\pi}}_f \mid \eta) \nonumber \\
    &\propto \left( \prod_{j=0}^{L_{\text{EM}}-1} \hat{\pi}_{f_j}^{N_j} \right) \left( \prod_{j=0}^{L_{\text{EM}}-1} \hat{\pi}_{f_j}^{\eta-1} \right) \nonumber \\
    &= \prod_{j=0}^{L_{\text{EM}}-1} \hat{\pi}_{f_j}^{N_j + \eta - 1} \sim \text{Dir}(\eta'_j), \nonumber
\end{align}
where the updated concentration parameters are
$\eta'_j = N_j + \eta$.
The MAP estimate for the $j$-th mixing coefficient is obtained by normalizing the mode of this posterior:
\begin{align}
    \hat{\pi}_{f_j} = \frac{\eta'_j - 1}{\sum_{i=0}^{L_{\text{EM}}-1} (\eta'_i - 1)} = \frac{N_j + \eta - 1}{\sum_i (N_i + \eta - 1)}. \label{eq:map_sol_final}
\end{align}
To enforce the sparsity-promoting property when $\eta < 1$, a thresholding operator is applied to ensure that the mixing-coefficient vector remains on the probability simplex
\begin{align}
    \hat{\pi}_{f_j} = \frac{\max(0, N_j + \eta - 1)}{\sum_i \max(0, N_i + \eta - 1)}. \nonumber
\end{align}
When $N_j$ is sufficiently small such that $N_j + \eta - 1 \leq 0$, the $j$-th state is effectively pruned from the model, since its mixing coefficient is driven to zero.

The Dirichlet concentration parameter $\eta$ therefore provides a principled mechanism for sparsity promotion by suppressing states with insufficient posterior support. Consequently, states that are activated only by a small number of observations tend to obtain negligible mixing coefficients and are automatically pruned during the MAP update. 

However, pruning based solely on the mixing coefficients does not eliminate all forms of model redundancy. In particular, multiple states may retain non-zero posterior probability while converging to nearly identical variance estimates. Although such states represent essentially the same noise regime, they are not removed by the Dirichlet prior because each state still receives sufficient statistical support. This issue is particularly relevant for HMM-EM. Through the BCJR forward-backward smoothing procedure, posterior probability is propagated across neighboring symbols using the estimated transition matrix. As a result, several similar states may continue to receive moderate posterior support and therefore remain above the Dirichlet pruning threshold.

To address this redundancy, the M-step can be appended with a variance-based state-merging step. After ordering the active states by variance, two adjacent states \(a\) and \(b\), with \(\hat{\sigma}_{f_a}^2\leq\hat{\sigma}_{f_b}^2\), are merged when
\[
\frac{\hat{\sigma}_{f_b}^2-\hat{\sigma}_{f_a}^2}
{\hat{\sigma}_{f_a}^2}<\tau_{\mathrm{merge}}.
\]
The merged mixing coefficient is \(\hat{\pi}_{f_a}+\hat{\pi}_{f_b}\), and the merged variance is their mixing-coefficient-weighted average. For HMM-M, the corresponding incoming and outgoing expected transition counts are aggregated, after which the redundant row and column are removed and the remaining transition matrix is row-normalized.

Consequently, the Dirichlet prior and the state-merging procedure play complementary roles. The Dirichlet prior removes weakly supported states, whereas state merging removes statistically redundant states. The merging threshold ultimately controls the trade-off between a compact model representation and the risk of underfitting the underlying noise distribution.

%% Complexity==============================================
\subsection{Complexity Analysis}
The E-step is the computational workhorse of the proposed EM algorithms because it evaluates the block likelihoods and infers the latent-state posteriors over all $N$ OFDM symbols. We therefore focus on its dominant multiplication and addition counts; the M-step, pruning and merging operations, divisions, exponential evaluations, and other nonlinear scalar operations are excluded from this comparison. The counts are stated as probability-domain multiplications and additions.

For SBS-EM, direct evaluation of the block likelihood in \eqref{eq:lik_em1} for one state and one OFDM symbol requires $M_w$ multiplications to form the squared magnitudes and $M_w-1$ multiplications to combine the scalar likelihoods. Multiplication by the state prior contributes one additional operation, giving $2M_wL_{\text{EM}}$ multiplications per OFDM symbol. Forming the denominator of \eqref{eq:pos_em1} requires $L_{\text{EM}}-1$ additions. Thus, one SBS-EM E-step over the frame requires
\begin{align}
    C_{\mathrm{SBS}}^{(\times)} = 2NM_wL_{\text{EM}}, \quad
    C_{\mathrm{SBS}}^{(+)} = N(L_{\text{EM}}-1). \nonumber
\end{align}

HMM-EM uses the same block likelihood but additionally performs the BCJR recursions over the $N-1$ transitions in the OFDM frame. For each transition, forming the branch metrics, executing the forward and backward recursions, and computing the state-pair posterior require $5L_{\text{EM}}^2$ multiplications in total. The summations in the forward recursion, backward recursion, and state marginalization require $3L_{\text{EM}}(L_{\text{EM}}-1)$ additions. Including the initial-state weighting and the state-posterior denominators, one HMM-EM E-step therefore requires
\begin{align}
    C_{\mathrm{HMM}}^{(\times)}
    &= NL_{\text{EM}}(2M_w-1)+L_{\text{EM}}
    +5(N-1)L_{\text{EM}}^2, \nonumber\\
    C_{\mathrm{HMM}}^{(+)}
    &= 3(N-1)L_{\text{EM}}(L_{\text{EM}}-1)
    +N(L_{\text{EM}}-1). \nonumber
\end{align}

The SBS-M and HMM-M variants retain the corresponding E-step structures but use the active state count at the beginning of iteration $i$, denoted by $L_{\text{act}}^{(i)}$. Thus, the multiplication and addition counts of the $i$-th SBS-M E-step are
\begin{align}
    C_{\mathrm{SBS-M}}^{(\times,i)}= 2NM_wL_{\text{act}}^{(i)}, \quad
    C_{\mathrm{SBS-M}}^{(+,i)} = N(L_{\text{act}}^{(i)}-1), \nonumber
\end{align}
whereas those of the $i$-th HMM-M E-step are
\begin{align}
    C_{\mathrm{HMM-M}}^{(\times,i)}
    &= NL_{\text{act}}^{(i)}(2M_w-1)+L_{\text{act}}^{(i)}
    +5(N-1)(L_{\text{act}}^{(i)})^2, \nonumber\\
    C_{\mathrm{HMM-M}}^{(+,i)}
    &= 3(N-1)L_{\text{act}}^{(i)}(L_{\text{act}}^{(i)}-1)
    +N(L_{\text{act}}^{(i)}-1). \nonumber
\end{align}
These expressions show that SBS-EM scales linearly with the number of states, whereas HMM-EM scales quadratically due to the BCJR state-transition processing. Consequently, reducing the active-state count through MAP pruning yields substantially larger computational savings for HMM-based estimation.

For numerical stability, the likelihood and BCJR calculations in this study are implemented in the log domain. Although this transformation replaces several multiplications by additions and uses the log-sum-exp operation for marginalization, it preserves the numbers of state--observation and state--transition metrics and therefore the relative dependence on $N$, $M_w$, and the state cardinality.

To verify these trends empirically, the mean wall-clock runtime of the log-domain implementations was measured over 100 independent trials using MATLAB R2024a on an Apple M2 Pro with 10 CPU cores and 16~GB of memory. Each trial used $M=64$, $M_w=16$, $N=1024$, an SNR of 5~dB, and Markov--Middleton parameters $A=0.05$, $\Lambda=1$, and $r=0.99$, with all algorithms executed for 20 EM iterations. The mean runtimes of SBS-10 and SBS-M were 19 and 13~ms, respectively, corresponding to a 30\% reduction. Similarly, HMM-10 and HMM-M required 1424 and 876~ms, respectively, corresponding to a 38\% reduction. At the final iteration, SBS-M and HMM-M retained an average of 5.13 and 5.50 active states, respectively, from the common 10-state initialization.

% Limitations==============================================
\subsection{Limitations and Assumptions of Block-EM}\label{sec:em_lim}
To obtain a tractable and computationally efficient approach for NSI estimation in OFDM systems using null tones, the proposed block-EM framework employs block-level aggregation. This aggregation preserves the dominant state-dependent power information, while enabling a simple frequency-domain likelihood and a practical receiver implementation. However, the resulting reduction in complexity is achieved through two key approximations.

Firstly, OFDM block processing compresses the ordered time-domain state sequence $\boldsymbol{s}^{(n)}$ into a state-count vector $\boldsymbol{c}^j$, which is subsequently represented by the count-weighted variance $\sigma_{f_j}^2$ in \eqref{eq:var_f}. As a result, the detailed temporal structure of the impulsive noise, such as the locations, the durations, or the ordering of the impulsive events within the OFDM symbol, is no longer preserved. Instead, it retains only the aggregate state-dependent power information represented by $\sigma_{f_j}^2$.

As the DFT size $M$ increases, different count vectors produce increasingly similar variance values, reducing the separability of the corresponding frequency-domain states. Furthermore, for a stationary finite-state Markov process with $r<1$, the CLT drives the frequency-domain IN distribution toward a single Gaussian distribution as $M\rightarrow\infty$. Consequently, the impulsive characteristics become progressively averaged out and the benefit of state-dependent frequency-domain likelihoods diminishes even when the frequency-domain NSI is perfectly known. We will further demonstrate this behavior in Sec.~\ref{sec:simu}.

The second approximation arises when the scalar frequency-domain model is used to construct a likelihood across multiple null tones. Conditioned on a particular time-domain state sequence, time-domain noise samples are independent Gaussian variables and therefore have a diagonal covariance matrix: $\mathbf{D}_{\boldsymbol{s}}=\mathrm{diag}(\boldsymbol{\sigma^2_{t}})$.
After applying the DFT, the frequency-domain noise vector becomes
\begin{align}
    \boldsymbol{W} \mid \boldsymbol{s}
    \sim
    \mathcal{CN}\left(\boldsymbol{W};
        \boldsymbol{0},
        \mathbf{F}\mathbf{D}_{\boldsymbol{s}}\mathbf{F}^{\mathsf{H}}
    \right). \nonumber
\end{align}
Although all subcarriers share the same marginal variance $\sigma^2_{f_j}$ associated with the count vector $\boldsymbol{c}^j$, the resulting covariance matrix is generally not diagonal. Consequently, different subcarriers may be statistically correlated whenever multiple time-domain states occur within the same OFDM symbol. 

The scalar GMM in \eqref{eq:fN} therefore provides an exact marginal distribution for each subcarrier. The product-form likelihood adopted by the EM estimators, however, treats the subcarriers as conditionally independent, effectively retaining only the diagonal entries of the full covariance matrix. This approximation avoids the computational burden of a full multivariate likelihood and enables a tractable receiver implementation.

As a result, some statistical efficiency may be lost because the estimator does not exploit cross-subcarrier correlations. Nevertheless, the dominant state-dependent power information is preserved for NSI estimation.

\section{Simulation Results}\label{sec:simu}
\iffigureson
\begin{figure*}[t]
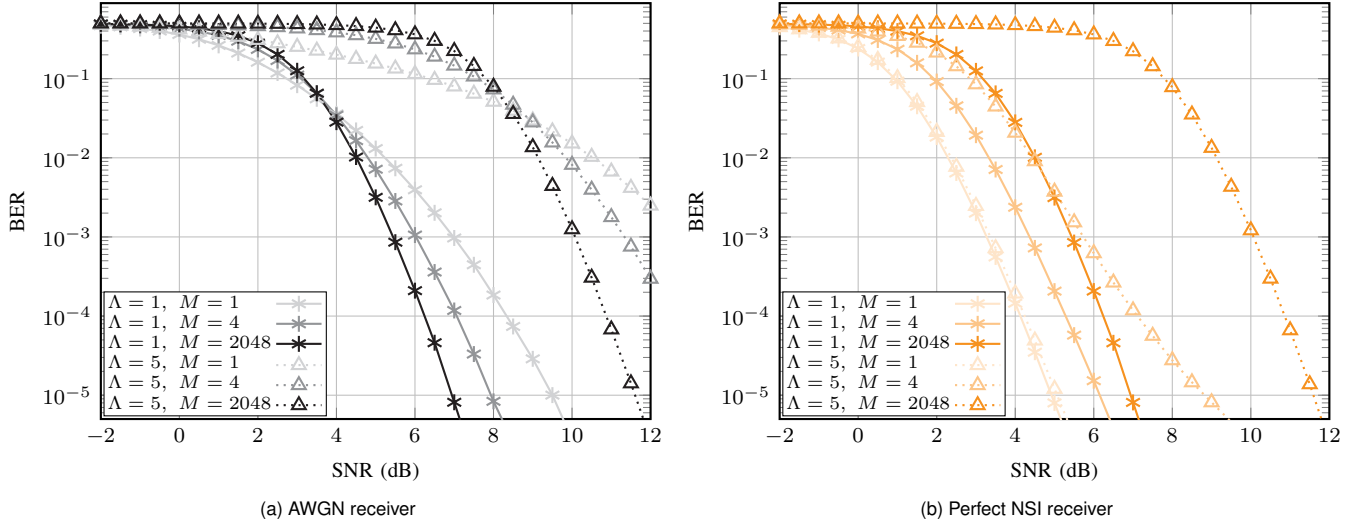

    \centering
    \subfloat[AWGN receiver \label{fig:ber_snr}]{%
        \centering
        \resizebox{1\columnwidth}{!}{\includegraphics{tikz/ber_snr.tikz}}
    }
    \subfloat[Perfect NSI receiver \label{fig:ber_snr_per}]{%
       \centering
       \resizebox{1\columnwidth}{!}{\includegraphics{tikz/ber_snr_per.tikz}}
    }
    \caption{BER performance versus SNR for (a) the conventional AWGN receiver and (b) the optimal receivers with perfect NSI knowledge with block sizes ($M$) of $1$, $4$, and $2048$ and impulsive-to-background noise ratio ($\Lambda$) of $1$ and $5$. The impulsive index and correlation parameter are fixed to $A=0.1$ and $r=0$, respectively.}
\end{figure*}
\fi

%% Simulation setup================================================
\subsection{Simulation Setup}
The performance of the proposed framework is evaluated using a coded OFDM system. We employ a standard rate-$1/2$ convolutional FEC scheme with a constraint length of $L_c=7$ and generator polynomials $(133, 171)_8$ in octal notation. Each frame contains a codeword of length $2^{17}$ and a random bit-interleaver of the same length as the codeword is applied. The interleaved bits are subsequently mapped to QPSK symbols before the OFDM modulation. At the receiver, the Viterbi algorithm is used for decoding to estimate the information bit sequence. We consider an OFDM system with a variable DFT size $M$ and a cyclic prefix of $M_{cp}=8$. 
The proposed algorithm is evaluated on a two-state ($L=2$) Markov--Middleton impulsive-noise channel with signal-to-noise ratio (SNR) defined with respect to the background noise variance as
$\text{SNR}=\mathbb{E}\{|X_{m,n}|^2\}/\sigma^2_{t_0}$.

\iffigureson
\begin{figure}
    \centering
    \resizebox{1\columnwidth}{!}{\includegraphics{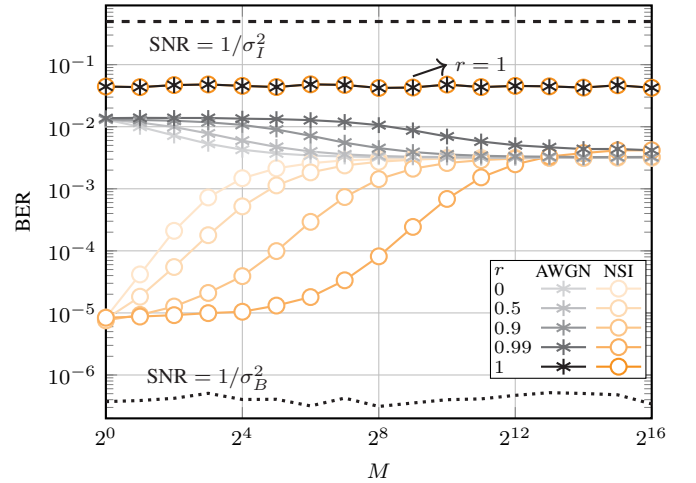}}
    \caption{BER over block size ($M$) for conventional (AWGN) and optimal (NSI) receivers with respect to different values of the correlation parameter $r$. The SNR for all simulations is fixed to $5$~dB. The impulsive index and impulse-to-background noise ratio are fixed to $A=0.1$ and $\Lambda=1$, respectively.}
    \label{fig:ber_nf_r}
\end{figure}
\fi

\iffigureson
\begin{figure*}[t]
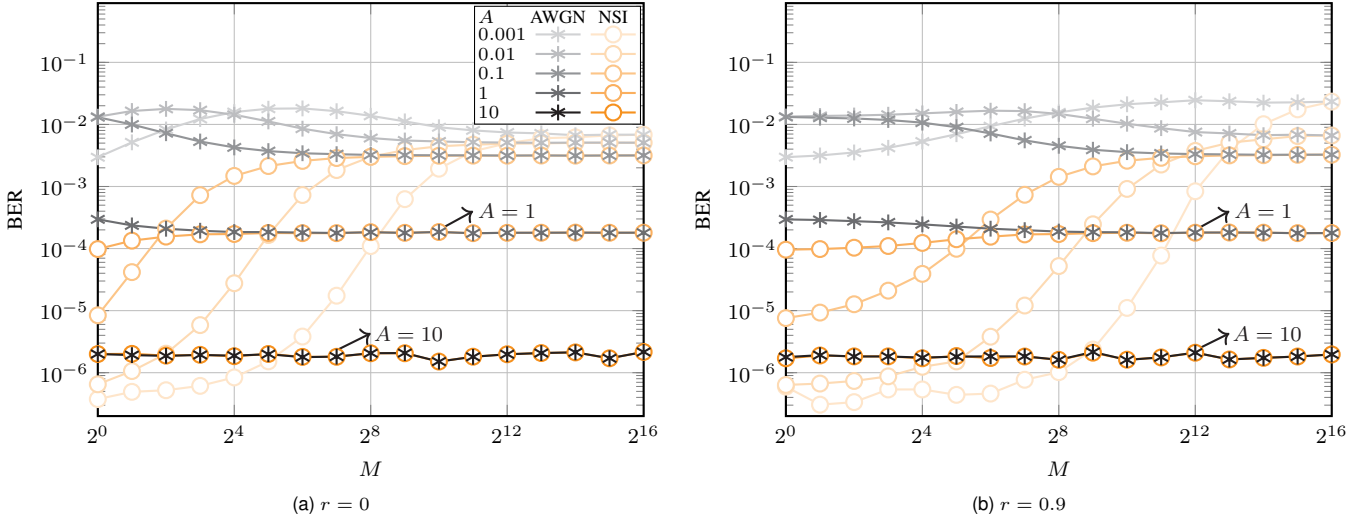

    \centering
    \subfloat[$r=0$ \label{fig:ber_nf_Ar0}]{
        \centering
        \resizebox{1\columnwidth}{!}{\includegraphics{tikz/ber_nf_A.tikz}}
    }
    \subfloat[$r=0.9$ \label{fig:ber_nf_Ar9}]{
       \centering
       \resizebox{1\columnwidth}{!}{\includegraphics{tikz/ber_nf_r09A.tikz}}
    }
    \caption{BER over block size ($M$) for conventional (AWGN) and optimal (NSI) receivers with respect to different values of the impulsive index $A$ for (a) $r=0$ and (b) $r=0.9$. The SNR for all simulations is fixed to $5$~dB, and the impulse-to-background noise ratio is set to $\Lambda=1$.}
\end{figure*}
\fi

%% Optimal Receiver===============================================
\subsection{Numerical Results: Optimal Receiver Performance}\label{sec:simu-opt}
We first evaluate the conventional AWGN receiver, which computes the likelihood in \eqref{eq:lik_con} by approximating the noise over the entire OFDM frame as a single Gaussian distribution with average variance $\hat{\sigma}_f^2 = \sum_{j=0}^{L_f-1} \pi_{f_j}\sigma^2_{f_j}$.

As shown in Fig.~\ref{fig:ber_snr}, the bit-error rate (BER) exhibits an SNR-dependent crossover for various block sizes $M$. At low SNR, a smaller $M$ minimizes the DFT's noise-smearing effect, confining dominant impulsive energy to fewer subcarriers. At high SNR, larger $M$ becomes advantageous, as the DFT spreads the impulsive energy over a wider bandwidth, providing a diversity-like gain that outweighs the error-spreading penalty~\cite{Suraweera04, Haring02, Nassar11}. As $\Lambda$ increases from $1$ to $5$, the channel impairment severely degrades performance; for the AWGN receiver at $M=2048$, the required SNR to achieve a $10^{-4}$ BER rises by approximately 5~dB.

Conversely, the optimal NSI-aware receiver using \eqref{eq:lik_per} demonstrates an inverse trend (Fig.~\ref{fig:ber_snr_per}): minimizing $M$ strictly improves BER. Because DFT block processing inherently averages noise across subcarriers, it smears highly resolved time-domain state information. Even with perfect NSI, this loss of instantaneous resolution degrades performance. Consequently, the optimal receiver maximizes its advantage at $M=1$, yielding a substantial 10~dB SNR gain at $10^{-4}$ BER for $\Lambda=5$. However, as $M$ increases, the CLT drives the smeared noise toward a single Gaussian distribution. The optimal receiver's performance rapidly converges to that of the conventional AWGN receiver, demonstrating that large block sizes inherently neutralize the benefits of state-dependent likelihoods.

Fig.~\ref{fig:ber_nf_r} illustrates BER versus $M$ under varying temporal correlation $r$ at a fixed $\text{SNR}=5$~dB. At $M=1$ and $r<1$, the optimal and AWGN receivers exhibit a massive performance gap (BERs of $10^{-5}$ vs. $10^{-2}$). As $M$ grows, both receivers converge to the CLT regime (e.g., at $M=64$ for $r=0$). However, higher temporal correlation preserves state stability against DFT smearing, allowing the optimal receiver to track the channel more accurately and maintain its advantage over much larger block sizes (converging at $M=4096$ for $r=0.9$ and $M=16384$ for $r=0.99$). In the extreme case of $r=1$ (fully static states per realization), the two receiver curves coincide and remain flat across all $M$, confirming that DFT spreading offers no benefit for static noise.

Finally, Figs.~\ref{fig:ber_nf_Ar0} and \ref{fig:ber_nf_Ar9} analyze the impulsive index $A$. Decreasing $A$ yields sparser but higher-power impulses. This sparsity magnifies the performance gain of the optimal NSI-aware receiver. When combined with high temporal memory ($r=0.9$ in Fig.~\ref{fig:ber_nf_Ar9}), the optimal receiver maintains an advantage of over three orders of magnitude at $M=256$, and remains 50 times better even at $M=4096$. In the extreme scenario for $A>1$, the model generates substantially more impulsive states than background states and asymptotically converges to a standard AWGN channel governed by the average impulsive power $\sigma^2_I$.

\begin{figure*}[t]
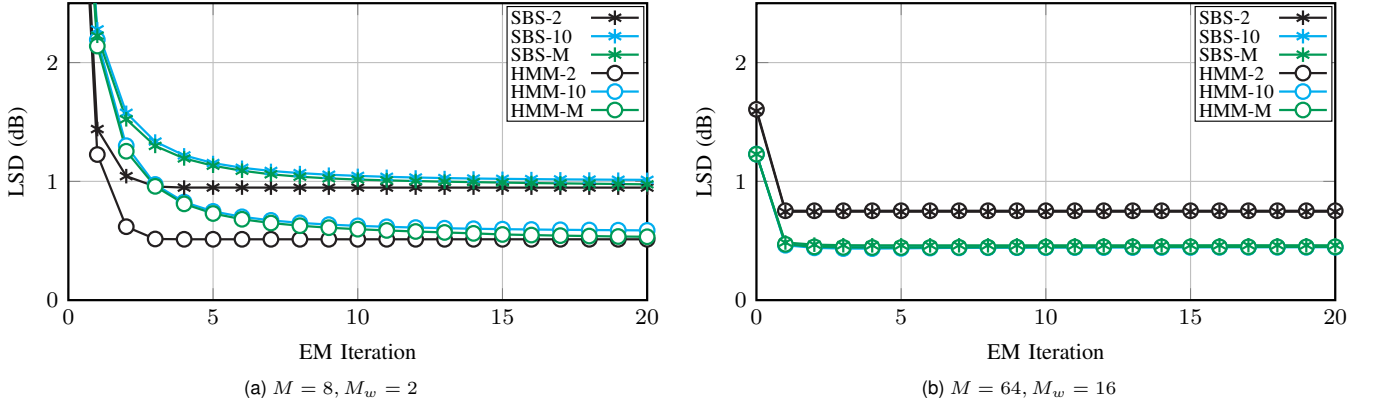

    \centering
    \subfloat[$M=8, M_w=2$ \label{fig:lsd_it_m8}]{
       \centering
       \resizebox{1\columnwidth}{!}{\includegraphics{tikz/lsd_it_M8.tikz}}
    }
    \subfloat[$M=64,M_w=16$ \label{fig:lsd_it_m64}]{
       \centering
       \resizebox{1\columnwidth}{!}{\includegraphics{tikz/lsd_it_M64.tikz}}
    }
    \caption{LSD performance versus EM iteration for various block-EM-based NSI estimation algorithms (SBS-EM, HMM-EM, and MAP-EM) for a DFT size of (a) $M=8$ and (b) $M=64$ with a null-tone overhead of $25$\%. The Markov--Middleton noise model parameters are set to $A=0.05$, $\Lambda = 1$, and $r=0.99$.}
\end{figure*}
\begin{figure*}[t]
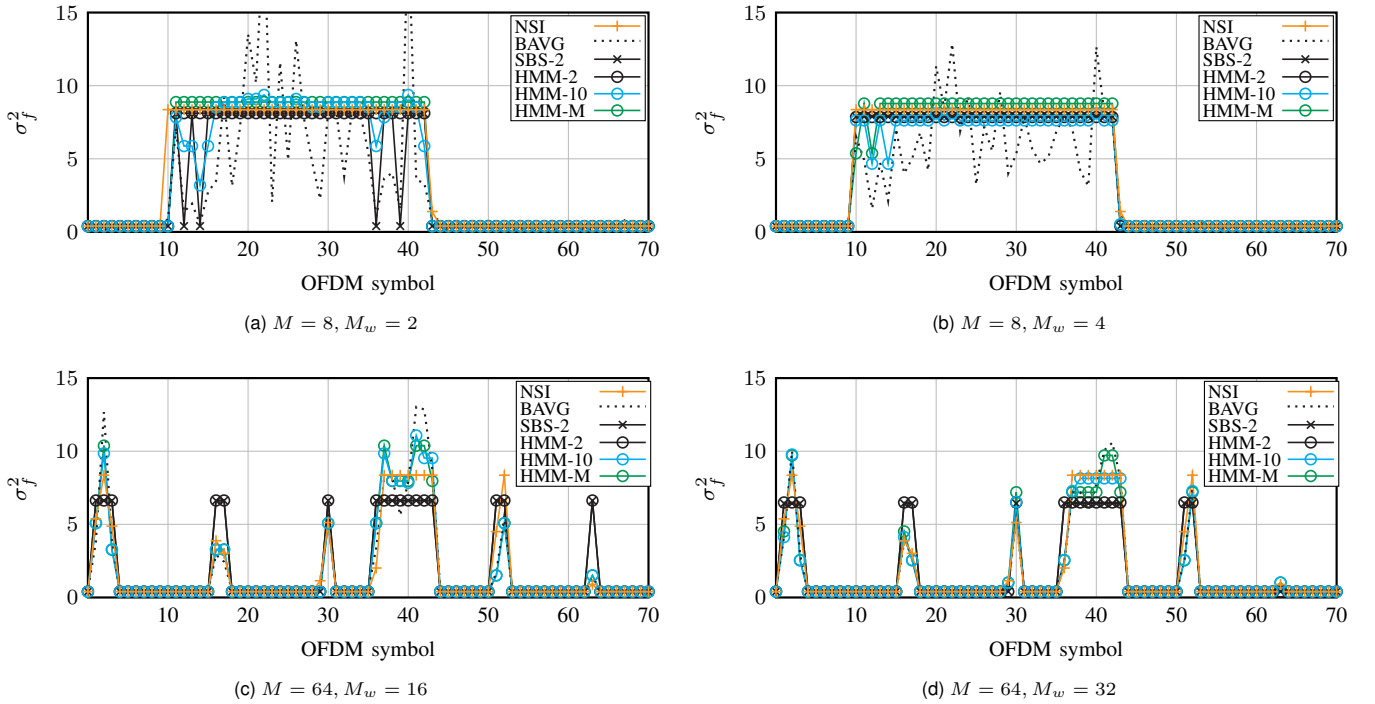

    \centering
    \subfloat[$M=8, M_w=2$ \label{fig:state_m8nn2}]{
       \centering
       \resizebox{1\columnwidth}{!}{\includegraphics{tikz/state_M8Nn2.tikz}}
    }
    \subfloat[$M=8, M_w=4$ \label{fig:state_m8nn4}]{
       \centering
       \resizebox{1\columnwidth}{!}{\includegraphics{tikz/state_M8Nn4.tikz}}
    }
    \vspace{0mm}
    \subfloat[$M=64, M_w=16$ \label{fig:state_m64nn16}]{
       \centering
       \resizebox{1\columnwidth}{!}{\includegraphics{tikz/state_M64Nn16.tikz}}
    }
    \subfloat[$M=64, M_w=32$ \label{fig:state_m64nn32}]{
       \centering
       \resizebox{1\columnwidth}{!}{\includegraphics{tikz/state_M64Nn32.tikz}}
    }
    \caption{Realization of variance estimation for various block-EM-based estimation algorithms (SBS-2, HMM-2, HMM-10, and MAP-HMM) in an OFDM system configured with DFT sizes $M\in\{8,64\}$ with available null-tone overheads of $25$\% and $50$\% under the Markov--Middleton noise model with $A=0.05$, $\Lambda = 1$, and $r=0.99$.  Solid orange lines represent the ground truth variance realizations, while dotted black lines correspond to the variance estimated using the per-OFDM-symbol averaging technique. }
\end{figure*}

%% Block-EM performance===========================================
\subsection{Numerical Results: Practical Block EM Receiver Performance}
In this subsection, the receiver performance of the proposed block-based EM frameworks is evaluated. The analysis focuses on scenarios in which the NSI is most critical--specifically, OFDM systems operating with relatively small DFT sizes under highly bursty IN conditions (i.e., low $A$ and high $r$), where accurate NSI estimation can yield substantial gains over an AWGN receiver. Following the sensitivity analysis in Sec.~\ref{sec:simu-opt}, the system is configured with $M \in \{8, 64\}$ and an IN channel with parameters $A=0.05$ and $r=0.99$.

The proposed finite-state symbol-by-symbol and HMM-based block-EM algorithms are denoted as {SBS-$L_{\text{EM}}$} and {HMM-$L_{\text{EM}}$}, where $L_{\text{EM}}$ represents the fixed number of states (e.g., SBS-2). For all algorithms, a range-based initialization strategy is employed for the noise variance: the state-dependent variances are linearly scaled between the $20$th and $100$th percentiles of the BAVG estimates. For the initial mixing coefficients $\boldsymbol{\pi}_f$, equal probabilities are assigned to ensure the EM algorithms with different state counts can freely redistribute weights during the iterations. Although further optimization of the initialization procedure is beyond the scope of this study, the adopted approach provides a representative and stable starting point for the state space.

The adaptive MAP-based EM variants are denoted as {SBS-M} and {HMM-M}, initialized with $L^0_{\text{act}}=10$ states and configured with a sparsity-promoting concentration parameter $\eta = 0.5$ and  $\tau_{\mathrm{merge}}=25\%$. These values are selected to balance state-pruning efficiency and fitting accuracy: $\eta=0.5$ remains in the sparsity-promoting Dirichlet regime without imposing an overly aggressive boundary condition, and the 25\% merging threshold removes redundant similar-variance components while avoiding the underfitting effect that can occur when the merging threshold is too large. 

The fidelity of the estimators is first evaluated using the log-spectral distance (LSD). The LSD provides a more balanced assessment of estimation accuracy across the diverse power levels inherent in the multi-state GMM compared to traditional linear error metrics. It is defined as
\begin{align}
    \text{LSD} = \sqrt{\mathbb{E} \left[\left( 10 \log_{10}(\sigma_f^2) - 10 \log_{10}(\hat{\sigma}_f^2) \right)^2\right]} \nonumber    
\end{align}
where $\mathbb{E}[\cdot]$ denotes the expectation over the ideal variances $\sigma_f^2$ and their corresponding estimates $\hat{\sigma}_f^2$ for each OFDM symbol. 

The selection of LSD is motivated by the high dynamic range characteristic of IN channels. In such environments, mean-square error (MSE) becomes dominated by estimation errors in high-variance states, obscuring the estimator’s behavior in low-power regimes. Although normalized MSE attempts to normalize the error, it becomes numerically unstable and overly pessimistic when the true variance approaches the noise floor, where even small absolute deviations lead to large relative errors. By operating in the logarithmic domain, the LSD provides a scale-invariant and robust measure that treats relative estimation errors uniformly across all states. Subsequently, the BER is used as the primary metric for evaluating end-to-end detection performance.

To evaluate the proposed frequency-domain block-EM variants in terms of both LSD and end-to-end BER, we consider the following reference receivers. The LSD comparison includes the NSI, AWGN, and BAVG benchmarks, whereas the BER comparison additionally includes the SBL receiver:
\begin{enumerate}
   \item \textit{NSI}: An ideal receiver with perfect NSI knowledge, serving as a performance upper bound.
   \item \textit{AWGN}: A conventional receiver using a single noise-variance estimate over the entire frame.
   \item \textit{BAVG}: An OFDM-symbol-level benchmark that estimates the noise variance separately for each OFDM symbol.
   \item \textit{SBL}: The state-of-the-art sparse Bayesian learning receiver proposed in \cite{Lin13}.
\end{enumerate}

Figs.~\ref{fig:lsd_it_m8} and \ref{fig:lsd_it_m64} illustrate the performance of the block-EM variants over successive iterations. At $M=8$ (Fig.~\ref{fig:lsd_it_m8}), the HMM-EM variants consistently outperform their SBS-EM counterparts by exploiting temporal correlation across consecutive OFDM symbols for robust state tracking. This advantage is evident in Fig.~\ref{fig:state_m8nn2}, where HMM-2 accurately tracks bursty impulses while SBS-2 frequently loses the state trajectory. Conversely, at $M=64$ (Fig.~\ref{fig:lsd_it_m64}), this temporal correlation weakens, diminishing the HMM advantage (Figs.~\ref{fig:state_m64nn16} and \ref{fig:state_m64nn32}). In this large-block regime, performance is driven by state-space cardinality rather than transition dynamics, allowing higher-order models (HMM-10 and HMM-M) to capture variance granularity significantly better than the 2-state baselines.

\begin{figure*}[t]
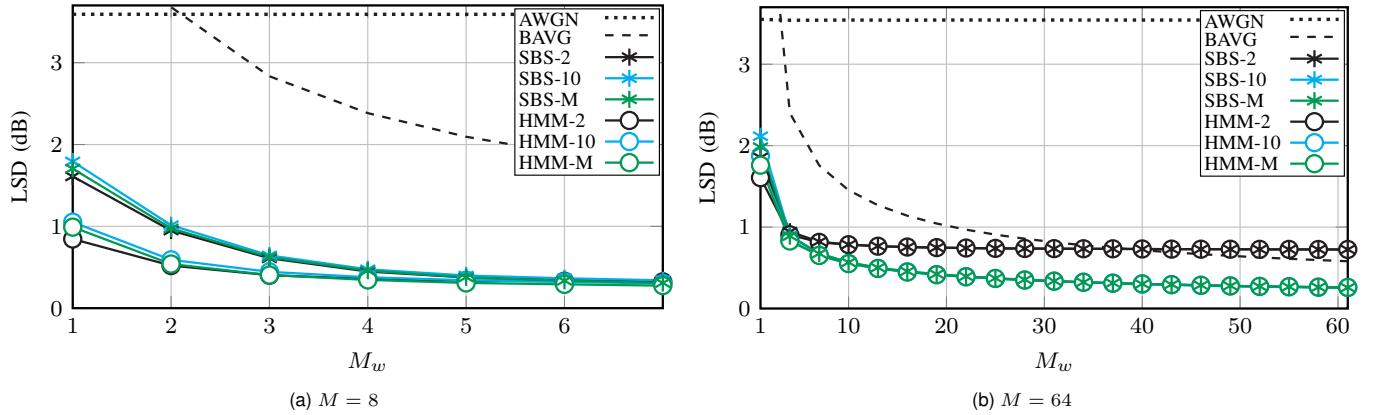

    \centering
    \subfloat[$M=8$ \label{fig:lsd_null_m8}]{
       \centering
       \resizebox{1\columnwidth}{!}{\includegraphics{tikz/lsd_null_M8.tikz}}
    }
    \subfloat[$M=64$ \label{fig:lsd_null_m64}]{
       \centering
       \resizebox{1\columnwidth}{!}{\includegraphics{tikz/lsd_null_M64.tikz}}
    }
    \caption{LSD performance versus the number of null tones of various block-EM-based NSI estimation algorithms (SBS-EM, HMM-EM, and MAP-EM) for a DFT size of (a) $M=8$ and (b) $M=64$. The Markov--Middleton noise model parameters are set to $A=0.05$, $\Lambda = 1$, and $r=0.99$. Dotted black curves represent the AWGN receiver, while dashed black curves denote the results obtained using the per-OFDM-symbol averaging methods.}
\end{figure*}

Figs.~\ref{fig:lsd_null_m8} and \ref{fig:lsd_null_m64} evaluate estimator performance against the null-tone count $M_w$. At $M=8$, HMM-EM provides massive gains at low overhead, yielding a 0.7 dB improvement over SBS-EM with just $M_w=1$ (12.5\% overhead). However, this gap vanishes as observation density increases to $M_w=7$ (87.5\% overhead), highlighting a fundamental design trade-off: high-overhead scenarios permit computationally simpler SBS estimators, whereas severely constrained systems require sophisticated HMM tracking. For $M=64$, 2-state models saturate early ($M_w > 16$), whereas higher-order and MAP-based variants continue to improve as additional observations become available.

Meanwhile, the baseline block-average (BAVG) estimator degrades severely in observation-constrained scenarios, performing worse than a standard AWGN receiver when $M_w<2$. BAVG's reliance on single-sample variance estimation induces high uncertainty and severe trajectory fluctuations (Fig.~\ref{fig:state_m8nn2}). In contrast, the proposed block-EM algorithms leverage the entire OFDM frame to update a finite state space, providing the statistical robustness necessary for stable estimation even under severe overhead constraints.

% MAP EM sensitivity analysis
Notably, the MAP-EM variants consistently match or exceed the fixed-order SBS-10 and HMM-10 models (Figs.~\ref{fig:lsd_null_m8} and \ref{fig:lsd_null_m64}). By employing a sparsity-promoting Dirichlet prior, MAP-EM dynamically prunes the initial 10-state space to a parsimonious configuration. 

\begin{figure*}[t]
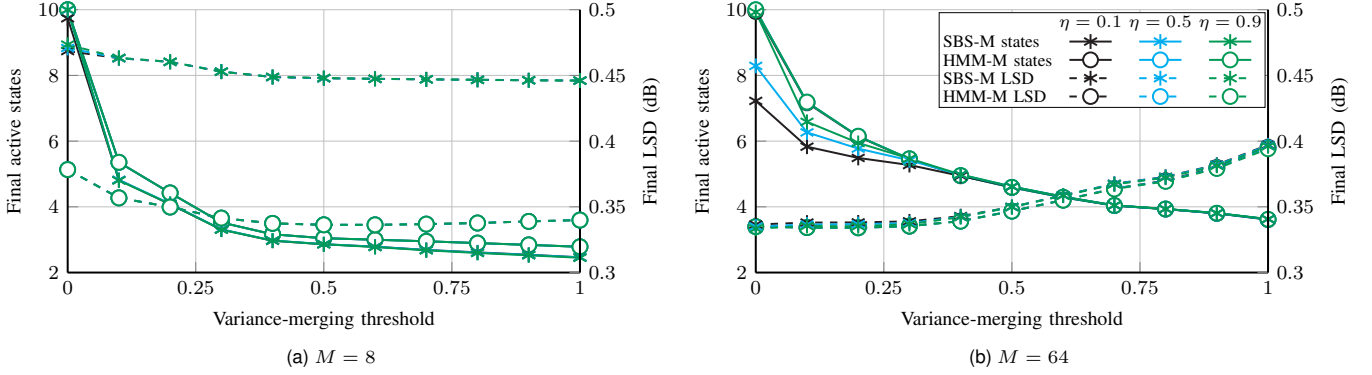

    \centering
    \subfloat[$M=8$ \label{fig:sens_merge_eta_m8}]{
       \centering
       \resizebox{1\columnwidth}{!}{\includegraphics{tikz/sens_merge_eta_M8.tikz}}
    }
    \subfloat[$M=64$ \label{fig:sens_merge_eta_m64}]{
       \centering
       \resizebox{1\columnwidth}{!}{\includegraphics{tikz/sens_merge_eta_M64.tikz}}
    }
    \caption{Sensitivity of the final active-state count and LSD to the Dirichlet concentration parameter $\eta$ and the variance-merging threshold for MAP-EM with DFT sizes of (a) $M=8$ and (b) $M=64$.}
    \label{fig:sens_merge_eta}
\end{figure*}

To evaluate the robustness of this adaptive pruning, Fig.~\ref{fig:sens_merge_eta} examines the effects of the Dirichlet concentration parameter $\eta$ and the variance-merging threshold. Without variance merging, SBS-M is more sensitive to $\eta$ than HMM-M. As $\eta$ increases from $0.1$ to $0.9$, the final SBS-M active-state count increases from about $7$ to $10$ for $M=64$, whereas HMM-M remains close to 10 states. 
This behavior occurs because smaller values of $\eta$ promote stronger sparsity. However, in HMM-M the BCJR forward-backward recursion propagates probability mass through the learned transition model, making individual states less sensitive to the choice of $\eta$ than in SBS-M. 
Once variance merging is introduced, the influence of $\eta$ becomes much weaker, and the merging threshold becomes the dominant factor controlling model size. 
For $M=8$, larger thresholds substantially reduce the active-state count without degrading LSD. For $M=64$, moderate merging ( $\tau_{\mathrm{merge}} <30\% $) successfully removes redundant states with minimal loss, whereas aggressive merging increases LSD because statistically distinct noise states are merged into a common component. 
The merging threshold therefore governs the trade-off between model compactness and underfitting.

\begin{figure*}[t]
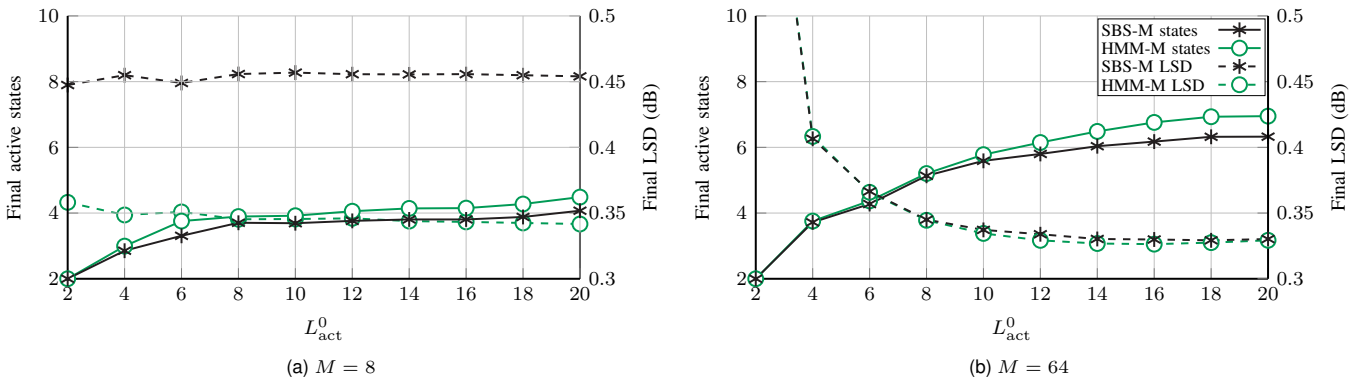

    \centering
    \subfloat[$M=8$ \label{fig:sens_lzero_m8}]{
       \centering
       \resizebox{1\columnwidth}{!}{\includegraphics{tikz/sens_L0_M8.tikz}}
    }
    \subfloat[$M=64$ \label{fig:sens_lzero_m64}]{
       \centering
       \resizebox{1\columnwidth}{!}{\includegraphics{tikz/sens_L0_M64.tikz}}
    }
    \caption{Sensitivity of the final active-state count and LSD to the initial active-state count $L^0_{\mathrm{act}}$ for MAP-EM with DFT sizes of (a) $M=8$ and (b) $M=64$.}
    \label{fig:sens_lzero}
\end{figure*}

Fig.~\ref{fig:sens_lzero} evaluates the sensitivity to the initial active-state count $L^0_{\mathrm{act}}$. 
For $M=8$, the final LSD is relatively insensitive to initialization once several initial states are available.
For $M=64$, the dependence is stronger because small initial state spaces cannot adequately represent the underlying frequency-domain noise distribution. They do not provide enough variance levels to approximate the underlying multi-state GMM. 
Increasing $L^0_{\mathrm{act}}$ improves LSD until the performance saturates around $L^0_{\mathrm{act}}\approx 12$--$16$, beyond which additional states provide little benefit. 
These results suggest that a moderate overestimate of the required state count followed by adaptive pruning is preferable to an overly restrictive initialization.

% as shown by the high LSD at $L^0_{\mathrm{act}}=2$ and the still elevated LSD at $L^0_{\mathrm{act}}=4$. Increasing $L^0_{\mathrm{act}}$ improves LSD until the performance reaches a plateau around $L^0_{\mathrm{act}}\approx 12$--$16$, after which additional initial states provide little benefit. These results support initializing MAP-EM with a moderate overestimate of the required state count, followed by adaptive pruning, particularly for the larger-DFT case considered here.

\begin{figure*}[t]
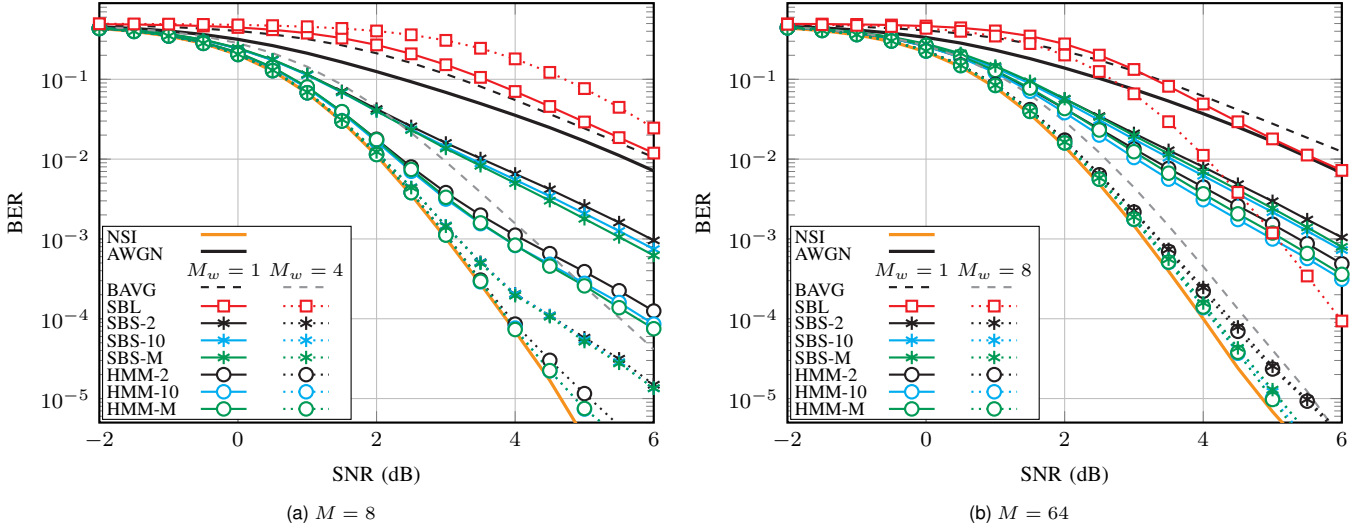

    \centering
    \subfloat[$M=8$ \label{fig:ber_snr_m8}]{
       \centering
       \resizebox{1\columnwidth}{!}{\includegraphics{tikz/ber_EM_M8.tikz}}
    }
    \subfloat[$M=64$ \label{fig:ber_snr_m64}]{
       \centering
       \resizebox{1\columnwidth}{!}{\includegraphics{tikz/ber_EM_M64.tikz}}
    }
    \caption{BER performance versus SNR for various block-EM-based NSI estimation algorithms (SBS-EM, HMM-EM, and MAP-EM) for a DFT size of (a) $M=8$ and (b) $M=64$. The Markov--Middleton noise model parameters are set to $A=0.05$, $\Lambda = 1$, and $r=0.99$. The performance is benchmarked against the NSI (orange curves), the conventional AWGN (black curves), and the per-OFDM-symbol averaging receivers (dashed black/gray curves).}
\end{figure*}

In Figs.~\ref{fig:ber_snr_m8} and \ref{fig:ber_snr_m64}, BER is used to evaluate the final detection performance across SNRs, highlighting the capability of the proposed framework to bridge the gap between conventional AWGN receivers and the ideal NSI performance bound. At low overhead ($M_w=1$), the HMM-based variants demonstrate a substantial advantage, with HMM-M achieving nearly a 2~dB gain over SBS-M at a BER of $10^{-3}$ for $M=8$ in Fig.~\ref{fig:ber_snr_m8}, underscoring the importance of temporal tracking via the BCJR algorithm. As the observation density increases to $M_w=4$ (50\% overhead), the HMM-EM variants closely approach the NSI performance, whereas the simplified SBS-EM solutions remain within approximately 0.5~dB of achieving the $10^{-4}$ BER target. 
For $M=64$ (Fig.~\ref{fig:ber_snr_m64}), the temporal correlation between consecutive symbols weakens, reducing the sensitivity of BER performance to HMM-based tracking. Notably, with a $12.5\%$ overhead ($M_w=8$), the block-EM receivers already approach the NSI bound.

Under the highly bursty IN conditions considered in Figs.~\ref{fig:ber_snr_m8} and \ref{fig:ber_snr_m64}, SBL yields a higher BER than all the proposed block-EM variants across the evaluated SNRs and null-tone configurations. 
This behavior is consistent with the difference between the underlying modeling assumptions. 
SBL reconstructs the time-domain IN by exploiting sample-level sparsity \cite{Lin13}. 
However, strong temporal correlation generates impulsive bursts that extend across multiple consecutive samples within an OFDM symbol, reducing the sparsity of the noise process and thereby weakening the assumptions on which SBL is based. Similar observations have also been reported in \cite{Berka26}.

\begin{figure}[t]
   \centering
   \resizebox{1\columnwidth}{!}{\includegraphics{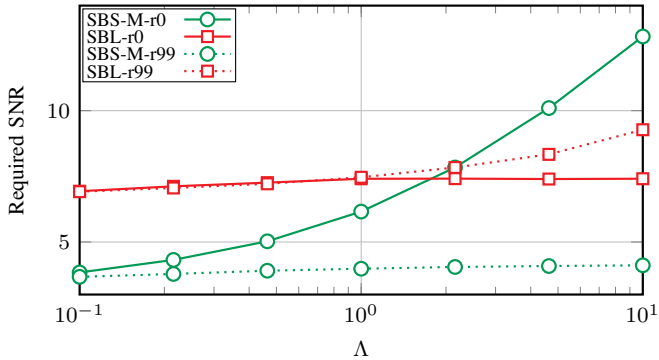}}

    \caption{Required SNR to achieve a BER of $10^{-4}$ versus $\Lambda$ for SBS-M and SBL for a DFT size of $M=64$ with $M_w=32$ null tones. The impulsive index is set to $A=0.05$.}
    \label{fig:required_snr_lambda}
\end{figure}

To identify the operating regions of SBL and the proposed block-EM framework, Fig.~\ref{fig:required_snr_lambda} compares the required SNR to achieve a BER of $10^{-4}$ as a function of $\Lambda$. 
For memoryless IN ($r=0$), SBL outperforms SBS-M only when the impulsive component is sufficiently stronger than the background noise, with a crossover at approximately $\Lambda=2$. 
In this regime, the impulses remain isolated and highly distinguishable, making sparse reconstruction particularly effective. 
In contrast, for strongly correlated IN ($r=0.99$), SBS-M requires a lower SNR across the entire evaluated range. This result is consistent with the information-theoretic analysis in \cite{Dario09} and \cite{chc25_1}, which highlight the value of exploiting temporal state information in bursty noise environments. 
These results indicate that SBL is best suited to sparse, memoryless, and highly distinctive IN, whereas the proposed block-EM is more robust when the impulsive component is less distinct or exhibits strong temporal correlation.

\section{Conclusions}\label{sec:conc}
This paper focuses on frequency-domain receiver implementation tailored for OFDM systems operating in bursty IN environments. The frequency-domain IN is modeled as a transformed GMM, whose model parameters are rigorously derived through a block transformation of the time-domain samples. An optimal receiver assuming perfect NSI is developed to serve as a benchmark. Moreover, a family of EM variants is designed based on a block likelihood formulation. 
The key practical system design trade-offs were highlighted through extensive evaluations. The results demonstrate that while a simplified two-state SBS-EM receiver is sufficient under high-overhead conditions, HMM-based tracking becomes essential when the null-tone overhead is limited. Additionally, higher-order EM variants provide improved performance in large-DFT systems, where the noise distribution exhibits greater granularity. Notably, the MAP-EM design achieves performance comparable to fixed 10-state models while offering substantial complexity reduction through its adaptive state-pruning capability.

Future research will investigate the optimization of initial conditions for EM-based NSI estimators, identified as a critical factor influencing the convergence behavior of low-order algorithms. Extending the proposed framework to semi-blind, pilot-aided OFDM systems represents a promising direction for further enhancing estimation reliability in highly dynamic environments. 
Moreover, future work should consider frequency-selective multipath channels with imperfect CSI. In this setting, the per-subcarrier channel response and the frequency-domain IN states become coupled estimation problems. Specifically, if the residual observations used for NSI estimation are formed using imperfect channel estimates, the residual contains both impulsive-noise realizations and channel-estimation errors. 
Consequently, the channel coefficients and the frequency-domain NSI may need to be estimated jointly or iteratively, similar in spirit to the approaches investigated in \cite{Berka26}.

% \section*{Acknowledgments}
% \input{tex/ACK}

\bibliographystyle{IEEEtran}
\bibliography{ref}

\vfill

\end{document}